\documentclass[a4paper,11pt]{article}
\pdfoutput=1 % if your are submitting a pdflatex (i.e. if you have
\usepackage{jheppub}

\usepackage{graphicx, epstopdf}

\usepackage[utf8]{inputenc}

\usepackage{amsmath, amsthm, amsfonts, amssymb}
\usepackage{mathtools, mathabx, bm, bbm, scalerel, xfrac}
\usepackage{empheq, physics, slashed, cancel}
\usepackage[usenames,dvipsnames]{xcolor}

\usepackage{hyperref}
\usepackage{color}

\def\ai{\alpha_i}
\def\aj{\alpha_j}
\def\gijq{g_{ijq}}
\def\li{\ln(-\alpha_i)}
\def\lipow#1{\ln^#1(-\alpha_i)}
\def\lj{\ln(-\alpha_j)}
\def\ljpow#1{\ln^#1(-\alpha_j)}
\def\lv{\ln(v)}
\def\lx{l_x}
\def\dilogx{\text{Li}_2(x)}
\def\msbar{$\overline{\text{MS}}$ }

\allowdisplaybreaks

\title{Subleading Effects in Soft-Gluon Emission at One-Loop in Massive QCD}

\author{Micha\l{} Czakon,}
\author{Kilian Erhard Minguez and}
\author{Felix Eschment}
\affiliation{Institut f\"ur Theoretische Teilchenphysik und Kosmologie, RWTH Aachen University,\\ D-52056 Aachen, Germany}
\emailAdd{mczakon@physik.rwth-aachen.de}
\emailAdd{kilian.erhard@rwth-aachen.de}
\emailAdd{felix.eschment@rwth-aachen.de}

\abstract{We provide the last missing ingredient necessary to approximate one-loop amplitudes in QCD with massive quarks in the limit of vanishing energy of a single gluon up to terms suppressed by this energy. Our main result is a soft operator acting in color and spin space that manipulates the momenta of the hard partons while keeping them on-shell and respecting momentum conservation. Additionally, we provide a complete expression for the subleading term of the expansion of an arbitrary tree-level amplitude in the limit where the momenta of a massless quark and a massless anti-quark of the same flavor become collinear. This limit is necessary to obtain the one-loop soft approximation whenever the process involves such a quark-anti-quark pair. Interestingly, the result involves a high-energy limit.}
\preprint{P3H-25-081, TTK-25-35}

\begin{document}
\maketitle
\flushbottom

\section{Introduction}

In a previous publication \cite{Czakon:2023tld}, we have studied the subleading single-soft-gluon asymptotics of one-loop amplitudes in QCD without massive quarks. There are, nevertheless, interesting settings where the masses of quarks play a role and should not be neglected. It is the purpose of the present work to provide a formula for this case as well. Whenever soft emissions from massive fermions are discussed, it is worth consulting the literature on QED. The most relevant works for us are Refs.~\cite{Engel:2021ccn,Engel:2023ifn,Engel:2023rxp} (see also the review \cite{Bailhache:2024mck}). Certainly, the gluon self-interaction makes the soft approximation substantially more complex in QCD, but gluon emissions from a single massive-quark line manifest some "abelian" properties which are given by the same expressions as in QED. It turns out that this is also the place where some interesting subtleties of factorization at subleading order are hidden. In particular, there are contributions due to hard virtual momenta, which are not contained in the one-loop amplitude for the process without gluon.

The soft asymptotics of amplitudes for the scattering of massive quarks only accompanied by a single gluon is entirely contained in a soft operator. The structure of the approximation is constrained by the Ward identity satisfied by the original amplitude and gauge-fixing independence of the amplitude with the gluon removed on which the operator is applied. This latter property requires the soft operator to at most change the color and spin of the quarks, while keeping their momenta on-shell and satisfying momentum conservation. We exploit these constraints to organize our result in a compact and transparent form.

The result is only complete if we also allow for massless quarks and gluons and not just massive quarks. Just as was the case in massless QCD, the formula for the soft operator, given as a sum over pairs of partons, is identical for every parton combination. At leading order of the soft expansion, this is a consequence of the well-known spin independence of the eikonal approximation. At subleading order, spin effects do play a role but amount to little-group transformations, which are the same for a massless quark and a massless gluon up to a normalization factor due to the different total spin.

Although we build upon Ref.~\cite{Czakon:2023tld}, we do not forget to mention some of the preceding studies of the topic most relevant to us in QED \cite{Gervais:2017yxv, Laenen:2020nrt}, QCD \cite{Bonocore:2015esa,Bonocore:2016awd}, SCET \cite{Larkoski:2014bxa,Beneke:2019oqx,Liu:2021mac} and gravity \cite{Beneke:2021umj,Beneke:2021aip,Beneke:2022pue}. All of them derive from the classic works of Low \cite{Low:1958sn}, Burnett and Kroll \cite{Burnett:1967km}, and Del Duca \cite{DelDuca:1990gz}. Interestingly, soft emissions at next-to-leading power still offer avenues for exploration even at tree-level as demonstrated by the very recent Refs.~\cite{vanBeekveld:2023gio,vanBeekveld:2023liw,Balsach:2023ema,Pal:2023vec,Pal:2024eyr}.

A side result of the study presented in Ref.~\cite{Czakon:2023tld} was the determination of the subleading behaviour of tree-level amplitudes in the two-parton collinear limit. In the case of splittings $\overset{\textbf{\fontsize{1pt}{1pt}\selectfont(--)}}{q} \to \overset{\textbf{\fontsize{2pt}{2pt}\selectfont(--)}}{q}g$  and $g \to gg$ the results were expressed in terms of universal factors and process-dependent amplitudes. A key element to obtain the expressions was the use of the subleading soft approximation for the final state gluons. As far as the splitting $g \to q\bar{q}$ is concerned, the structure of the result was determined, but one term was left unexpressed through universal factors and process-dependent amplitudes unlike what would be expected from a factorization-based result. The reason for this shortcoming was the lack of an expression for the subleading soft asymptotics due to a (anti-)quark. Since subleading tree-level collinear asymptotics are a key ingredient for the one-loop subleading-soft approximation, it is desirable to have all results provided in the most general and intuitively appealing form possible. We present the last missing piece needed for the splitting $g \to q\bar{q}$ in the present publication.

The paper is organized as follows. In the next section, we define the problem and quote a minimum of formulae to keep the publication reasonably self-contained. Afterwards, we discuss the master integrals present in the result. Although these are known for a long time, we reproduce the most relevant expressions for completeness. Our main result is presented in Section~\ref{sec:S1}. There, we not only provide the formula for the soft operator, but also discuss pole cancellation, and, most importantly, the intricacies of the derivation. Appendix~\ref{app:gqq} on the subleading collinear limit for $g \to q\bar{q}$ closes the publication.

\section{Definitions}

We consider the process
\begin{equation} \label{eq:process}
    0 \to a_1(p_1, \sigma_1, c_1) + \dots + a_n(p_n, \sigma_n, c_n) + g(q, \sigma_{n+1}, c_{n+1}) \; , \qquad a_i \in \{q, \bar{q}, g\} \; ,
\end{equation}
where $a_i$, $p_i$, $\sigma_i$, $c_i$ are the respective flavors, momenta, polarizations, and colors of the partons. We do not make a distinction between (anti-)quarks of different flavor as QCD interactions are flavor independent. The singled-out gluon is assumed to be soft, i.e.\ $q^0 \ll | p_i^0 |$. We allow for massive quarks, $p_i^2 \neq 0$, $a_i \in \{q, \bar{q}\}$ for at least one\footnote{The result is valid beyond QCD and flavor-changing interactions might allow for amplitudes with a single massive quark.} $i$. The process may include any number of color singlets, which we omit in the dicussion for the sake of simplicity of notation. The reduced process without the soft gluon is
\begin{equation} \label{eq:reduced}
    0 \to a_1(p_1', \sigma_1', c_1') + \dots + a_n(p_n', \sigma_n', c_n') \; .
\end{equation}
Amplitudes for a process with $m$ partons, with $m = n+1$ or $m = n$, are expanded in the bare strong coupling constant $g_s^B$ as follows.
\begin{equation} \label{eq:ME}
    M_m \equiv \big( g_s^B )^{m-2} \bigg[ M_m^{(0)} + \frac{\big( \mu^2 e^{\gamma_{\text{E}}} \big)^{-\epsilon} \alpha^B_s}{(4\pi)^{1-\epsilon}} \, M_m^{(1)} + \order{\big( \alpha_s^B \big)^2} \bigg] \; , \qquad \alpha^B_s \equiv \frac{(g^B_s)^2}{4\pi} \; .
\end{equation}
The results are presented in dimensional regularization with spacetime dimension $d \equiv 4 - 2\epsilon$. The strong coupling constant is left unrenormalized, but mass countererms are included as they are needed for gauge-fixing independence of the amplitudes. Of course, LSZ normalization factors are also included. The massive quarks are considered as decoupled, i.e.\ there are no heavy quark loops. 

The structure of the soft approximation at one-loop,
\begin{equation} \label{eq:OneLoopLBK}
\begin{split}
    &\ket*{M_g^{(1)}(\{p_i\},q)} \approx \mathbf{S}^{(0)}(\{p_i\},\{p_i'\},q) \, \ket*{M^{(1)}(\{p'_i\})} + \mathbf{S}^{(1)}(\{p_i\},\{p_i'\},q) \, \ket*{M^{(0)}(\{p'_i\})} \\[.2cm]
    &\qquad + \int_0^1 \dd{x} \bigg( \sum_{i, \, m_i = 0} \mathbf{J}_i^{(1)}(x,p'_i,q) \ket*{H^{(0)}_{g,i}(x,\{p'_i\},q)} + \sum_{\substack{i, \, a_i = g}} \mathbf{\tilde{J}}_i^{(1)}(x,p'_i,q) \ket*{H^{(0)}_{\bar{q},i}(x,\{p'_i\},q)} \bigg) \\
    &\qquad + \sum_{\substack{i \neq j \\ m_i = 0 \\ m_j = 0}} \sum_{\substack{\tilde{a}_i \neq a_i \\ \tilde{a}_j \neq a_j}} \mathbf{\tilde{S}}^{(1)}_{a_i a_j \, \leftarrow \, \tilde{a}_i \tilde{a}_j, \, ij}(p'_i,p'_j,q) \, \ket*{M^{(0)}(\{p'_i\}) \, \big|{\substack{a_i \, \to \, \tilde{a}_i \\ a_j \, \to  \, \tilde{a}_j}}} \; ,
\end{split}
\end{equation}
has been determined in Ref.~\cite{Czakon:2023tld}. $\ket*{M_g^{(1)}}$ is the one-loop amplitude for process \eqref{eq:process} considered as a vector in color-and-spin space, while $\ket*{M^{(0)}}$, $\ket*{M^{(1)}}$ are the tree-level and one-loop amplitudes for process \eqref{eq:reduced}. The definitions of the modified $(n+1)$-parton tree-level amplitudes in the collinear limit, $\ket*{H^{(0)}_{g,i}}$ and $\ket*{H^{(0)}_{\bar{q},i}}$, as well as the definitions of the operators $\mathbf{J}_i^{(1)}$, $\mathbf{\tilde{J}}_i^{(1)}$ and $\mathbf{\tilde{S}}^{(1)}$ can be found in Ref.~\cite{Czakon:2023tld}. The extension of the one-loop soft operator $\mathbf{S}^{(1)}$ to the massive case is the subject of the present publication. The tree-level soft operator $\mathbf{S}^{(0)}$ is given by\footnote{At the subleading level in the soft expansion, it is not necessary to make a distinction between $p_i$ and $p_i'$. On the one hand, this results in a lack of strict power-counting in the soft expansion parameter. On the other hand, however, the expressions are more compact. The latter property is particularly welcome in the massive case.}
\begin{equation} \label{eq:LBK}
    \mathbf{P}_g(\sigma,c) \, \mathbf{S}^{(0)} = - \sum_i \mathbf{T}_i^c \otimes \mathbf{S}^{(0)}_i \; , \qquad
    \mathbf{S}^{(0)}_i = \frac{p_i \cdot \epsilon^*}{p_i \cdot q} \Big( 1 + \sum_j \big( p_i - p_i' \big) \cdot \partial_j \Big)  +  \frac{1}{2 p_i \cdot q} F_{\mu\nu} \big( J^{\mu\nu}_i - \mathbf{K}^{\mu\nu}_i \big) \; ,
\end{equation}
where $\mathbf{P}_g(\sigma,c)$ is a surjection operator from $(n+1)$-parton to $n$-parton color-and-spin space that removes the gluon state by projecting out the polarization $\sigma$ and color $c$. The respective gluon polarization vector is $\epsilon^* \equiv \epsilon^*(\bm{q},\sigma)$. The tensor-product notation separates operators acting in color space from operators acting in spin space. $\mathbf{T}_i^c$ is a standard color operator acting on the color state of parton $i$, and $F_{\mu\nu}$ is the abelian field-strength tensor in the momentum representation,
\begin{equation}
    F_{\mu\nu} = i \big( q_\mu \epsilon^*_\nu - q_\nu \epsilon^*_\mu \big) \; .
\end{equation}
Finally, $J^{\mu\nu}_i$ is the four-dimensional angular-momentum tensor operator,
\begin{equation}
    J^{\mu\nu}_i \equiv i \big( p_i^\mu \partial_i^\nu - p_i^\nu \partial_i^\mu \big) \; , \qquad \partial_i^\mu \equiv \pdv*{p_{i\mu}} \; ,
\end{equation}
and $\mathbf{K}^{\mu\nu}_i$ is a spin-space operator acting on the polarization state of parton $i$
\begin{equation}
    \mathbf{K}_i^{\mu\nu} \ket{\dots,\sigma_i',\dots} \equiv \sum_{\sigma_i} K^{\mu\nu}_{a_i, \, \sigma_i\sigma_i'} \ket{\dots,\sigma_i,\dots} \; ,
\end{equation}
defined through its matrix elements $K^{\mu\nu}_{a,\sigma \sigma'}$,
\begin{equation} \label{eq:K}
\begin{aligned}
    &\sum_{\sigma'} K^{\mu\nu}_{q,\sigma\sigma'} \, \bar{u}(\bm{p}_i,\sigma') \equiv J_i^{\mu\nu} \, \bar{u}(\bm{p}_i,\sigma) - \frac{1}{2} \bar{u}(\bm{p}_i,\sigma) \, \sigma^{\mu\nu} \; , \\[.2cm]
    &\sum_{\sigma'} K^{\mu\nu}_{g,\sigma\sigma'} \, \epsilon^*_\alpha(\bm{p}_i,\sigma') \equiv \Big( J_i^{\mu\nu} \, g_{\alpha\beta} + i \big( \delta^\mu_\alpha \delta^\nu_\beta - \delta^\nu_\alpha \delta^\mu_\beta \big) \Big) \, \epsilon^{\ast\beta}(\bm{p}_i,\sigma) + \text{terms proportional to } p_{i\,\alpha} \; .
\end{aligned}
\end{equation}
The matrix elements of $\mathbf{K}_i$ can be obtained with the help of Lorentz transformations of spinors and polarization vectors. For example
\begin{equation}
    S(\Lambda) \, u(\bm{p},\sigma) = \sum_{\sigma'} \mathcal{D}^{(1/2)}_{\sigma'\sigma}(R) \, u(\bm{p}',\sigma') \; , \qquad
    \Lambda = e^\omega \; , \qquad
    S = e^{- \frac{i}{4} \omega_{\mu\nu} \sigma^{\mu\nu}} \; , \qquad
    p' = \Lambda \, p \; ,
\end{equation}
where $\mathcal{D}^{(1/2)}_{\sigma'\sigma}(R)$ is the Wigner rotation matrix for the little-group rotation $R$, which depends on the conventions used in the definition of the spinor. For more details, see Ref.~\cite{Czakon:2023tld}.

\section{Master integrals} \label{sec:masters}

The one-loop soft operator $\mathbf{S}^{(1)}$ is given in Section~\ref{sec:S1} in terms of six master integrals. 
One of those is the standard one-loop integral $A_0(m^2)$ which vanishes for $m = 0$ and is otherwise given by
\begin{equation}
    A_0(m^2) = \mu^{2\epsilon}e^{\epsilon \gamma_{\text{E}}} \int \frac{d^d l}{i \pi^{d/2}} \frac{1}{[l^2-m^2]} = \frac{1}{\epsilon} + 1 + \ln\Big(\frac{\mu^2}{m^2}\Big) + \order{\epsilon} \; .    
\end{equation}
For convenience, we also define
\begin{equation}
    \bigg( -\frac{\mu^2 s_{ij}}{s_{iq}s_{jq}} \bigg)^{\epsilon} M_{0i} \equiv \frac{A_0(m_i^2)}{m_i^2} \; ,
\end{equation}
with
\begin{equation}
    s_{ij} \equiv 2 p_i \cdot p_j + i 0^+ \; , \qquad
    s_{iq} \equiv 2 p_i \cdot q + i 0^+ \; , \qquad
    s_{jq} \equiv 2 p_j \cdot q + i 0^+ \; .
\end{equation}
The remaining five master integrals,
\begin{equation} \label{eq:M123}
\begin{gathered}
    M_{1i} \equiv (p_i \cdot q) \, I_{1110} \; , \qquad
    M_{1j} \equiv (p_j \cdot q) \, I_{1101} \; , \\[.2cm]
    M_{2j} \equiv \frac{1}{2} (p_i \cdot p_j) \, I_{1011} \; , \qquad
    M_{2i} \equiv \frac{1}{2} (p_i \cdot p_j) \, I_{0111} \; , \\[.2cm]
    M_{3} \equiv (p_i \cdot q)(p_j \cdot q)  \, I_{1111} \; ,
\end{gathered}
\end{equation}
are defined in terms of the generic integral $I_{\alpha_1\alpha_2\alpha_3\alpha_4}$,
\begin{equation} \label{eq:GenericInt}
    \bigg( -\frac{\mu^2 s_{ij}}{s_{iq}s_{jq}} \bigg)^{\epsilon} I_{\alpha_1\alpha_2\alpha_3\alpha_4} \equiv 
    \big( \mu^2 e^{\gamma_{\text{E}}} \big)^\epsilon \int \frac{d^d l}{i \pi^{d/2}} \frac{1}{[l^2]^{\alpha_1} [(l + q)^2]^{\alpha_2} [p_i \cdot (l + q)]^{\alpha_3} [-p_j \cdot l]^{\alpha_4}} \; .
\end{equation}
Each expression in the square brackets on the r.h.s.\ of Eq.~\eqref{eq:GenericInt} contains an implicit $+i0^+$. The integrals have been calculated in Ref.~\cite{Bierenbaum:2011gg} with exact dependence on the spacetime dimension $d$. For most applications at subleading order in the soft expansion, only a Laurent expansion to $\order{\epsilon^0}$ is needed. Assuming $m_i \neq 0$ and $m_j \neq 0$, there is
\begin{equation}
\begin{aligned}
    M_{1k} &= \frac{1}{2\epsilon^2} + \frac{\ln(-\alpha_k)}{2\epsilon} + \frac{1}{4} \ln^2(-\alpha_k) + \frac{5\pi^2}{24} + \order{\epsilon} \; , \\[.2cm]
    M_{2k} &= \frac{1}{v} \bigg( - \frac{l_x}{2\epsilon} + \frac{1}{2} \ln(-\alpha_k) \, \lx - \lv \, \lx - \frac{1}{4} \lx^2 - \text{Li}_2(x) + \frac{\pi^2}{6} \bigg) + \order{\epsilon} \; , \\[.2cm]
    M_3 &= \frac{1}{\epsilon^2} + \frac{\li + \lj}{2\epsilon} + \frac{1}{2} \li \lj - \frac{1}{4} \lx^2 - \frac{7\pi^2}{12} + \order{\epsilon} \; ,
\end{aligned}
\end{equation}
with
\begin{equation}
\begin{gathered}
    \ai \equiv \frac{m_i^2 s_{jq}}{s_{ij} s_{iq}} \; , \qquad
    \aj \equiv \frac{m_j^2 s_{iq}}{s_{ij} s_{jq}} \; , \qquad
    v \equiv \sqrt{1-4 \ai \aj} = \sqrt{1-\frac{m_i^2 \, m_j^2}{(p_i \cdot p_j)^2}} \; , \\[.2cm]
    x \equiv \frac{1-v}{1+v} \; , \qquad
    \lx \equiv \ln(x) + 2 \pi i \, \theta(s_{ij}) \; .
\end{gathered}
\end{equation}
The analytic continuation of $\ln(x)$ to timelike kinematics, $s_{ij} > 0$, follows from the relation
\begin{equation}
    x = \ai \aj (1+x)^2 = \bigg( \frac{m_i m_j (1+x)}{-s_{ij}} \bigg)^2 \; ,
\end{equation}
which implies
\begin{equation}
    \lx = \li + \lj +2 \ln(1+x) = 2 \ln(m_i m_j (1+x)) - 2 \ln(-s_{ij}) \; .
\end{equation}
Since $|x| \leq 1$, $1+x$ does not cross the branch cut of the logarithm. On the one hand, expressing $\lx$ in terms of $\li$, $\lj$ and $\ln(1+x)$ allows for easy analytic continuation and provides direct access to mass singularities. On the other hand, however, expressions for the integrals are more compact in terms of $\lx$.

In case $m_j = 0$ and $m_i \neq 0$, $M_{1i}$ remains unchanged, but there is
\begin{equation}
\begin{aligned}
    M_{1j} &= 0 \; , \qquad M_{2j} = -M_{1i} \; , \\[.2cm]
    M_{2i} &= \frac{1}{2\epsilon^2} - \frac{\li}{2\epsilon} + \frac{1}{4} \li^2 + \frac{3\pi^2}{8} + \order{\epsilon} \; , \\[.2cm]
    M_3 &= \frac{3}{2\epsilon^2} + \frac{\li}{2\epsilon} - \frac{1}{4} \li^2 - \frac{5\pi^2}{24} + \order{\epsilon} \; .
\end{aligned}
\end{equation}
In case $m_i = 0$ and $m_j \neq 0$, the master integrals are obtained by swapping $i$ and $j$.
Finally, if $m_i = m_j = 0$ then
\begin{equation}
    M_{1k} = M_{2k} = 0 \; , \qquad M_3 = \frac{2}{\epsilon^2} + \frac{\pi^2}{6} + \order{\epsilon} \; .
\end{equation}
It has been proved in Ref.~\cite{Bierenbaum:2011gg} that the leading term of the soft expansion is regular at vanishing masses. It follows from Eq.~\eqref{eq:g1ij} that $M_3+M_{2j}$ and $M_{2i}+M_{1j}$ are regular at $m_j = 0$. Similarly, $M_3+M_{2i}$ and $M_{2j}+M_{1i}$ are regular at $m_i = 0$. This implies that the sum of the five master integrals
\begin{equation}
    M_3+M_{2i}+M_{2j}+M_{1i}+M_{1j} 
\end{equation}
is regular at both $m_i = 0$ and $m_j = 0$.

\section{The one-loop soft operator} \label{sec:S1}

The one-loop soft operator can be decomposed into non-abelian (NA) and abelian (A) contributions. It is a sum over partons $i$ and $j$ with a summand that contains the subleading soft behavior due to $i$.
\begin{equation} \label{eq:S1}
\begin{split}
    \mathbf{P}_g(\sigma,c) \, \mathbf{S}^{(1)} &= \sum_{i \neq j} \bigg( - \frac{\mu^2 s_{ij}}{s_{iq} s_{jq}} \bigg)^\epsilon \, \bigg[ 2 i f^{abc} \mathbf{T}_i^a \mathbf{T}_j^b \otimes \mathbf{S}^{(\text{NA})}_{ij} + \mathbf{T}_i^c \, \mathbf{T}_i \cdot \mathbf{T}_j \, S^{(\text{A})}_{ij} \bigg] \\[.2cm]
    &\quad + C_F \sum_i \mathbf{T}_i^c \, \frac{2\epsilon(1-\epsilon)(1+2\epsilon)}{1-2\epsilon} \, \frac{A_0(m_i^2)}{m_i^2} \, \bigg( g^{\mu\nu} - \frac{2 p_i^\mu p_i^\nu}{m_i^2} \bigg) \, \frac{F_{\mu\rho}}{s_{iq}} \, \mathbf{K}_{i \, \nu}{}^\rho \; ,
\end{split}
\end{equation}
\begin{equation}
\begin{split}
    \mathbf{S}^{(\text{NA})}_{ij} &=  - g^{(1)}_{ij} \, \mathbf{S}^{(0)}_{i} + \epsilon \, S^{(\text{A})}_{ij} + a_i \, \bigg( g^{\mu\nu} - \frac{2 p_i^\mu p_i^\nu}{m_i^2} \bigg) \, \frac{F_{\mu\rho}}{s_{iq}} \, \mathbf{K}_{i \, \nu}{}^\rho \label{eq:SNAij} \\[.2cm]
    &\quad + \bigg( b_{ij} \, \aj \, g^{\mu\nu} + c_{ij} \, \aj \, \frac{2 p_j^\mu p_j^\nu}{m_j^2} + b_{ji} \, \frac{2 p_j^\mu p_i^\nu}{s_{ij}} + c_{ji} \, \frac{2 p_i^\mu p_j^\nu}{s_{ij}} \bigg) \, \frac{F_{\mu\rho}}{s_{iq}} \big( J_i - \mathbf{K}_i \big)_\nu{}^\rho \\[.2cm]
    &\quad + \frac{b_{ji} + c_{ji} + 2 \aj \big( b_{ij} + c_{ij} \big)}{v^2} \, \frac{2 p_i^\mu p_j^\nu F_{\mu\nu}}{s_{ij} s_{iq}} \, \frac{2 p_i^\rho p_j^\sigma}{s_{ij}} \big( J_i - \mathbf{K}_i \big)_{\rho\sigma} \; ,
\end{split}
\end{equation}
\begin{align}
    - g^{(1)}_{ij} &= \frac{M_{1i} + M_{1j} + ( 1 - 2 \aj ) \, M_{2i} + ( 1 - 2 \ai ) \, M_{2j} + ( 1 - 2 \ai - 2 \aj) \, M_3}{2 \gijq} \; , \label{eq:g1ij} \\[.4cm]
    S^{(\text{A})}_{ij} &= - \frac{8}{v^2} \bigg( \frac{2\epsilon }{1-2\epsilon} \, M_{1i} + \big( v^2 - 1 \big) \, M_{2j} \bigg) \, \frac{p_i^\mu p_j^\nu \, i F_{\mu\nu}}{s_{ij} s_{iq}} \; , \quad
    a_i = \frac{(1-\epsilon)(1 - 2 \epsilon^2) \, M_{0i} - 2\epsilon \, M_{1i}}{1-2\epsilon} \; , \label{eq:ai} \\[.4cm]
    b_{ij} &= \frac{- M_{1i} - M_{1j} + (1 - 2 \ai) ( M_{2i} - M_{2j} ) + M_3}{\gijq} \; , \\[.4cm]
    c_{ij} &= \bigg[ \Big( 1 - 2 \aj - (1-2\epsilon) \big( 1 - 2 \ai \big) \Big) \, M_{1i} - \Big( 1 - 2 \ai + 2 \gijq - (1-2\epsilon) \big( 1 - 2 \aj + 2 \gijq \big) \Big) \, M_{1j} \nonumber \\[.2cm]
    &\quad + \Big( 2 (1-2\epsilon) \big( v^2 - \gijq \big) + 2\epsilon \big( 1 - 2 \ai \big)^2 \Big) \, M_{2i} + \Big( 2 \gijq - 2 (1-\epsilon) \, v^2 \Big) \, M_{2j} \nonumber \\[.2cm]
    &\quad + \Big( 1 - 2 \ai - (1-2\epsilon) \big( 1 - 2 \aj \big) \Big) \, M_3 \bigg] \frac{1}{2 (1-2\epsilon) \, \gijq^2} \; , \\[.4cm]
    \gijq & = \frac{4 \det(G)}{s_{ij} s_{iq} s_{jq}} = 1-\ai-\aj \; ,
\end{align}
where $G$ is the Gram matrix of $p_i,p_j$ and $q$.
Several comments are in order.
\begin{enumerate}

\item The soft operator is not UV renormalized as implied by the definition of the matrix elements \eqref{eq:ME}. The UV renormalized version is given by
\begin{equation}
    \mathbf{S}^{(1)} - \frac{\beta_0}{2\epsilon} \mathbf{S}^{(0)} \; , \qquad \beta_0 = \frac{11}{3} C_A - \frac{4}{3} T_F n_l \; ,
\end{equation}
with $n_l$ the number of massless quark flavors as we have decoupled heavy quarks.

\item The coefficient $g^{(1)}_{ij}$ determines the leading behavior of the soft operator. It has been originally obtained in Ref.~\cite{Bierenbaum:2011gg}\footnote{The minus sign in front of $g^{(1)}_{ij}$ in Eq.~\eqref{eq:SNAij} is necessary to compensate for the opposite sign of the strong coupling constant assumed in Ref.~\cite{Bierenbaum:2011gg}.}.

\item The subleading behavior of the soft operator when $i$ is a massless quark is identical to that when $i$ is a gluon, up to the difference in the value of the spin operators $\mathbf{K}_i$, which are then diagonal in the helicity basis.

\item The abelian contribution is a new feature of the massive case not present in massless QCD. Our result for $S^{(\text{A})}_{ij}$ agrees with the corresponding expression in QED provided in Ref.~\cite{Engel:2021ccn}. The term in the second line of Eq.~\eqref{eq:S1} does not affect unpolarized matrix elements, but is necessary at the amplitude level. It was first determined for QED in Refs.~\cite{Engel:2022kde,Kollatzsch:2022bqa}. This term can in principle be combined with $S^{(\text{A})}_{ij}$ using the following relation valid  by virtue of color conservation for summands independent of $j$.
\begin{equation}
    \sum_{i \neq j} \mathbf{T}_i^c \, \mathbf{T}_i \cdot \mathbf{T}_j = - C_F \sum_i \mathbf{T}_i^c \; .
\end{equation}

\item Contributions proportional to the master integral $A_0$ originate in the hard-region of the soft expansion, see Section~\ref{sec:derivation}. We keep them in the soft operator $\mathbf{S}^{(1)}$ in order to maintain the structural minimality of Eq.~\eqref{eq:OneLoopLBK}. In an analysis based on effective field theory, these contributions would be separated as they correspond to the propagation of hard modes rather than soft modes\footnote{See for example Ref.~\cite{Engel:2023ifn} for a discussion within QED.}.

\item The term in the second line of Eq.~\eqref{eq:S1} as well as the third term in Eq.~\eqref{eq:SNAij} could as well be expressed through $J_i - \mathbf{K}_i$ because
\begin{equation}
    \bigg( g^{\mu\nu} - \frac{2 p_i^\mu p_i^\nu}{m_i^2} \bigg) \, \frac{F_{\mu\rho}}{s_{iq}} \, J_{i \, \nu}{}^\rho = 0 \; .
\end{equation}

\item If $m_i = 0$ then the last term in Eq.~\eqref{eq:SNAij} can be combined with the term proportional to $b_{ji}$. Indeed, equations of motion imply in this case
\begin{equation}
    \frac{2 p_i^\mu p_j^\nu F_{\mu\nu}}{s_{ij} s_{iq}} \, \frac{2 p_i^\rho p_j^\sigma}{s_{ij}} \big( J_i - \mathbf{K}_i \big)_{\rho\sigma} = - \frac{2 p_j^\mu p_i^\nu}{s_{ij}} \frac{F_{\mu\rho}}{s_{iq}} \,  \big( J_i - \mathbf{K}_i \big)_\nu{}^\rho \; .
\end{equation}
If $i$ is a gluon, then transversality of the process-dependent amplitude $\ket*{M^{(0)}}$ implies furthermore
\begin{equation}
    p_j^\mu p_i^\nu F_{\mu\rho} \,  \big( J_i - \mathbf{K}_i \big)_\nu{}^\rho  = 0 \; .
\end{equation}

\item If $m_i \neq 0$ then the coefficient of the last term in Eq.~\eqref{eq:SNAij} is related to $b_{ij}$ and $c_{ij}$ due to momentum-conservation constraints. Indeed, the soft operator must satisfy
\begin{equation}
    \mathbf{S}^{(1)} \big|_{\substack{\text{momentum} \\ \text{derivatives}}} \ket*{f(P')} = 0 \; , \qquad P' \equiv \sum_{i} p'_i \; ,
\end{equation}
where $\ket*{f}$ is an arbitrary function of the sum of the momenta $p'_i$ of the particles in the reduced process. This translates to
\begin{equation} \label{eq:MomentumConservation}
\Big[ \mathbf{S}^{(\text{NA})}_{ij} - \mathbf{S}^{(\text{NA})}_{ji} \Big]_{\substack{\text{momentum} \\ \text{derivatives}}} \ket*{f(P')} = 0 \; .
\end{equation}
The first term in Eq.~\eqref{eq:SNAij} satisfies this constraint because
\begin{equation}
    \mathbf{S}^{(0)}_i \big|_{\substack{\text{momentum} \\ \text{derivatives}}} \, \ket*{f(P')} = - \epsilon^* \cdot \pdv{P'} \ket*{f(P')} \; ,
\end{equation}
which is independent of $i$, and because the coefficient, $-g^{(1)}_{ij}$, of $\mathbf{S}^{(0)}_i$ is symmetric in $i$ and $j$, while the color factor is anti-symmetric. For the remaining terms containing $J_i$ one obtains the relation specified in the last term of Eq.~\eqref{eq:SNAij}.

\end{enumerate}

\subsection{Spurious-pole cancellation} \label{sec:S1Poles}

The pole terms of the soft operator \eqref{eq:S1} depend on the mass configuration.
\begin{enumerate}
    \item $m_i \neq 0$ and $m_j \neq 0$
    \begin{equation} \label{eq:S1Poles}
    \begin{aligned}
        \mathbf{S}^{(\text{NA})}_{ij} &= \frac{1}{\epsilon} \Bigg[ \frac{1}{\epsilon} + \frac{1}{2} \bigg( \li + \lj - \frac{\lx}{v} \bigg) \Bigg] \, \mathbf{S}^{(0)}_i + \order{\epsilon^0} \; , \\[.2cm]
        S^{(\text{A})}_{ij} &= \frac{1}{\epsilon} \Bigg[ - \frac{8}{v^2} \bigg( 1 + \frac{2 \ai \aj \, \lx}{v} \bigg) \, \frac{p_i^\mu p_j^\nu }{s_{ij} s_{iq}} \, i F_{\mu\nu} \Bigg] + \order{\epsilon^0} \; .
    \end{aligned}
    \end{equation}
    \item $m_i \neq 0$ and $m_j = 0$
    \begin{equation} \label{eq:S1Poles0j}
    \begin{aligned}
        \mathbf{S}^{(\text{NA})}_{ij} &= \frac{1}{\epsilon} \Bigg[ \frac{1}{\epsilon} \, \mathbf{S}^{(0)}_i + 4 \, \frac{p_j^\mu p_j^\nu}{s_{ij} s_{jq}} \, F_{\mu\rho} \big( J_i - \mathbf{K}_i \big)_\nu{}^\rho \Bigg] + \order{\epsilon^0} \; , \\[.2cm]
        S^{(\text{A})}_{ij} &= \frac{1}{\epsilon} \Bigg[ - 8 \, \frac{p_i^\mu p_j^\nu }{s_{ij} s_{iq}} \, i F_{\mu\nu} \Bigg] + \order{\epsilon^0} \; .
    \end{aligned}
    \end{equation}
    \item $m_i = 0$ and $m_j \neq 0$
    \begin{equation} \label{eq:S1Poles0i}
    \begin{aligned}
        \mathbf{S}^{(\text{NA})}_{ij} &= \frac{1}{\epsilon} \Bigg[ \frac{1}{\epsilon} \, \mathbf{S}^{(0)}_i + 4 \, \frac{p_i^\mu p_j^\nu - p_j^\mu p_i^\nu}{s_{ij} s_{iq}} \, F_{\mu\rho} \big( J_i - \mathbf{K}_i \big)_\nu{}^\rho  \Bigg] + \order{\epsilon^0} \; , \\[.2cm]
        S^{(\text{A})}_{ij} &= 0 \; ,
    \end{aligned}
    \end{equation}
    \item $m_i = 0$ and $m_j = 0$
    \begin{align}
        \mathbf{S}^{(\text{NA})}_{ij} &= \frac{1}{\epsilon} \Bigg[ \frac{1}{\epsilon} \, \mathbf{S}^{(0)}_i + 4 \, \frac{p_i^\mu p_j^\nu - p_j^\mu p_i^\nu}{s_{ij} s_{iq}} \, F_{\mu\rho} \big( J_i - \mathbf{K}_i \big)_\nu{}^\rho + 4 \, \frac{p_j^\mu p_j^\nu}{s_{ij} s_{jq}} \, F_{\mu\rho} \big( J_i - \mathbf{K}_i \big)_\nu{}^\rho \Bigg] + \order{\epsilon^0} \; , \nonumber \\[.2cm]
        S^{(\text{A})}_{ij} &= 0 \; , \label{eq:S1Poles0ij}
    \end{align}
\end{enumerate}
Using the known singularities of massive one-loop gauge-theory amplitudes \cite{Catani:2000ef}, one may obtain the pole terms of the soft approximation\footnote{This is a generalization of Eq.~(4.78) in Ref.~\cite{Czakon:2023tld}.},
\begin{equation} \label{eq:SoftApproxPoles}
\begin{split}
    \mathbf{P}_g(\sigma,c) \, &\ket*{M_g^{(0)}} \approx \mathbf{P}_g(\sigma,c) \, \mathbf{S}^{(0)} \, \ket*{M^{(1)}} \\[.2cm]
    & + \sum_{i \neq j} 2 i f^{abc} \mathbf{T}_i^a \mathbf{T}_j^b \otimes \bigg( - \frac{\mu^2 s_{ij}}{s_{iq} s_{jq}} \bigg)^\epsilon \, \frac{1}{\epsilon} \Bigg[ \frac{1}{\epsilon} + \frac{1}{2} \bigg( \li + \lj - \frac{\lx}{v} \bigg) \Bigg] \, \mathbf{S}^{(0)}_i \ket*{M^{(0)}} \\[.2cm]
    & + \sum_{i \neq j} \mathbf{T}_i^c \, \mathbf{T}_i \cdot \mathbf{T}_j \, \frac{1}{\epsilon} \Bigg[ - \frac{8}{v^2} \bigg( 1 + \frac{2 \ai \aj \, \lx}{v} \bigg) \, \frac{p_i^\mu p_j^\nu }{s_{ij} s_{iq}} \, i F_{\mu\nu} \Bigg] \ket*{M^{(0)}} + \order{\epsilon^0} \; .
\end{split}
\end{equation}
The pole terms of the soft operator match this formula for $m_i \neq 0$ and $m_j \neq 0$ as confirmed by Eq.~\eqref{eq:S1Poles}. The r.h.s.\ of Eq.~\eqref{eq:SoftApproxPoles} is regular if either $m_i = 0$ or $m_j = 0$,
\begin{equation} \label{eq:SoftApproxPoles0}
\begin{split}
    \mathbf{P}_g(\sigma,c) \, \ket*{M_g^{(0)}} &\approx \mathbf{P}_g(\sigma,c) \, \mathbf{S}^{(0)} \, \ket*{M^{(1)}} \\[.2cm]
    & + \sum_{i \neq j} 2 i f^{abc} \mathbf{T}_i^a \mathbf{T}_j^b \otimes \bigg( - \frac{\mu^2 s_{ij}}{s_{iq} s_{jq}} \bigg)^\epsilon \frac{1}{\epsilon} \Bigg[ \frac{1}{\epsilon} \Bigg] \, \mathbf{S}^{(0)}_i \ket*{M^{(0)}} \\[.2cm]
    & + \sum_{i \neq j} \mathbf{T}_i^c \, \mathbf{T}_i \cdot \mathbf{T}_j \, \frac{1}{\epsilon} \Bigg[ - 8 \, \frac{p_i^\mu p_j^\nu }{s_{ij} s_{iq}} \, i F_{\mu\nu} \Bigg] \ket*{M^{(0)}} + \order{\epsilon^0} \; .
\end{split}
\end{equation}
Eqs.~\eqref{eq:S1Poles0j}, \eqref{eq:S1Poles0i} and \eqref{eq:S1Poles0ij} clearly differ from this result. However, if either $i$ or $j$ is massless then it is necessary to include jet-operator and flavor-off-diagonal soft contributions in Eq.~\eqref{eq:OneLoopLBK}. In particular, if $j$ is massless then\footnote{This result is obtained by summing Eqs.~(4.80)~and~(4.82) from Ref.~\cite{Czakon:2023tld} and exchanging $i$ and $j$. The term in the first line in Eq.~\eqref{eq:JetPoles} contains $\big( J_i - \mathbf{K}_i \big)$ instead of $\big( J_j - \mathbf{K}_j \big)$ which it should have contained according to Eq.~(4.80) of Ref.~\cite{Czakon:2023tld}. The reason for the difference is that symmetry w.r.t.\ the exchange of the summation indices $i$ and $j$ has been used in Ref.~\cite{Czakon:2023tld}. In the present case, $i$ and $j$ cannot be exchanged, since only one of the partons is massless, and there is no jet-operator contribution for massive quarks.}
\begin{equation} \label{eq:JetPoles}
\begin{split}
  \mathbf{P}_g(\sigma,c) \, &\int_0^1 \dd{x} \mathbf{J}_j^{(1)} \ket*{H^{(0)}_{g,j}} = 
    \sum_{i \neq j} 2 i f^{abc} \mathbf{T}_i^a \mathbf{T}_j^b \otimes \frac{1}{\epsilon} \, \Bigg[ - 4 \frac{p_j^\mu p_j^\nu}{s_{ij} s_{jq}} F_{\mu\rho} \, \big( J_i - \mathbf{K}_i \big)_{\nu}{}^{\rho} \Bigg] \, \ket*{M^{(0)}} \\[.2cm]
    &+ \sum_{i \neq j} 2 i f^{abc} \mathbf{T}_i^a \mathbf{T}_j^b \otimes \frac{1}{\epsilon} \, \Bigg[ - 4 \frac{p_i^\mu p_j^\nu - p_j^\mu p_i^\nu}{s_{ij} s_{jq}} F_{\mu\rho} \, \big( J_j - \mathbf{K}_j \big)_{\nu}{}^{\rho} \Bigg] \, \ket*{M^{(0)}} \\[.2cm]
    &+ \sum_{i \neq j} \mathbf{T}_j^c \, \mathbf{T}_j \cdot \mathbf{T}_i \, \frac{1}{\epsilon} \Bigg[ - 8 \frac{p_j^\mu p_j^\nu}{s_{ij} s_{jq}} iF_{\mu\nu} \Bigg] \, \ket*{M^{(0)}} + \order{\epsilon^0}\\[.2cm]
    &+ \text{terms canceled by an $\mathbf{\tilde{S}}^{(1)}$ contribution if $j$ is a (anti-)quark} \; .
\end{split}
\end{equation}
The sum of each of Eqs.~\eqref{eq:S1Poles0j}, \eqref{eq:S1Poles0i}, \eqref{eq:S1Poles0ij} and Eq.~\eqref{eq:JetPoles} for either $j$ or $i$ or both, depending on the mass configuration, matches the pole terms of Eq.~\eqref{eq:SoftApproxPoles0}.

\subsection{Finite contributions}

As in the case of pole terms in Section~\ref{sec:S1Poles}, the finite part of the soft operator depends on the mass configuration. Below we list the values of the finite parts of the coefficients defined in Eqs.~\eqref{eq:S1}~and~\eqref{eq:SNAij}.
\begin{enumerate}
    \item $m_i \neq 0$ and $m_j \neq 0$
    \begin{align} \label{eq:S1Finite}
        - g^{(1)}_{ij} &= \frac{1}{v} \Bigg[ \bigg( \frac{(1-2\aj)\li+(1-2\ai)\lj }{4\gijq} - \lv - \frac{1}{4} \lx \bigg) \, \lx - \dilogx + \frac{\pi^2}{6} \Bigg] \nonumber \\[.2cm]
        &\quad + \frac{\big( \li-\lj \big)^2 + \lx^2 + 4\pi^2}{8\gijq} + \frac{1}{2} \li \lj - \frac{1}{4} \lx^2 - \frac{7\pi^2}{12} \; , \\[.4cm]
        \epsilon S^{(\text{A})}_{ij} &= - \frac{8}{v^2} \Bigg[ 1 + \frac{2 \ai \aj \, \lx}{v} \Bigg] \, \frac{p_i^\mu p_j^\nu \, i F_{\mu\nu}}{s_{ij} s_{iq}} \; , \\[.4cm]
        S^{(\text{A})}_{ij} &= - \frac{8}{v^2} \Bigg[ 2 + \li + \frac{\ai \aj}{v} \bigg( \big( 4 \lv - 2 \lj + \lx \big) \, \lx + 4 \dilogx - \frac{2\pi^2}{3} \bigg) \Bigg] \, \frac{p_i^\mu p_j^\nu \, i F_{\mu\nu}}{s_{ij} s_{iq}} \; , \\[.2cm]
        a_i &= - 2 \li - 2 \ln(-s_{ij}) + 2 \ln(-s_{jq}) \; , \\[.4cm]
        b_{ij} &= \frac{1}{4\gijq} \Bigg[ \bigg( \lj - \li + \frac{1 - 2 \ai}{v} \, 2 \lx \bigg) \big( \li - \lj \big) - \lx^2 - 4\pi^2 \Bigg] \; , \\[.4cm]
        c_{ij} &= \frac{1}{2\gijq} \Bigg[ \frac{(\ai-\aj) \Big( \big( \li - \lj \big)^2 + \lx^2 + 4\pi^2 \Big) + 2v \big( \li - \lj \big) \, \lx}{2\gijq} \nonumber \\
        &\qquad\qquad + \big( \lj - \li - 2 \big( 1 - 2 \ai \big) \big) \frac{\lx}{v} + 2 \big( \li - \lj \big) \Bigg] \; .
    \end{align}
    \item $m_i \neq 0$ and $m_j = 0$
    \begin{equation}
    \begin{aligned} \label{eq:S1Finite0j}
        - g^{(1)}_{ij} &= \frac{\ai \lipow2 + (1+5\ai) \frac{\pi^2}{6}}{2(1-\ai)} \; , \\[.2cm]
        \epsilon S^{(\text{A})}_{ij} &= -8 \, \frac{p_i^\mu p_j^\nu \, i F_{\mu\nu}}{s_{ij} s_{iq}} \; , \qquad
        S^{(\text{A})}_{ij} = \big( -16-8 \li \big) \, \frac{p_i^\mu p_j^\nu \, i F_{\mu\nu}}{s_{ij} s_{iq}} \; , \\[.2cm]
        a_i &= - 2 \li - 2 \ln(-s_{ij}) + 2 \ln(-s_{jq}) \; , \qquad
        b_{ij} = \frac{-\ai \lipow2 + (1-7\ai) \frac{\pi^2}{6}}{1-\ai} \; , \\[.4cm]
        c_{ij} &= 4 + \frac{2\ai \li}{1-\ai} + \frac{\ai \big( \lipow2 + \pi^2 \big)}{(1-\ai)^2} \; .
    \end{aligned}
    \end{equation}
    \item $m_i = 0$ and $m_j \neq 0$
    \begin{equation}
    \begin{aligned} \label{eq:S1Finite0i}
        - g^{(1)}_{ij} &= \frac{\aj \ljpow2 + (1+5\aj) \frac{\pi^2}{6}}{2(1-\aj)} \; , \qquad
        \epsilon S^{(\text{A})}_{ij} = 0 \; , \qquad
        S^{(\text{A})}_{ij} = 0 \; , \qquad
        a_i = 0 \; , \\[.4cm]
        b_{ij} &= - \frac{\ljpow2+\pi^2}{1-\aj} \; , \qquad
        c_{ij} = - \frac{2\lj}{1-\aj} - \frac{\aj \big( \ljpow2+\pi^2 \big)}{(1-\aj)^2} \; .
    \end{aligned}
    \end{equation}
    \item $m_i = 0$ and $m_j = 0$
    \begin{equation} \label{eq:S1Finite0ij}
        - g^{(1)}_{ij} = \frac{\pi^2}{12} \; , \qquad
        \epsilon S^{(\text{A})}_{ij} = 0 \; , \qquad
        S^{(\text{A})}_{ij} = 0 \; , \qquad
        a_i = 0 \; , \qquad
        b_{ij} = \frac{\pi^2}{6} \; , \qquad
        c_{ij} = 4 \; .
    \end{equation}
\end{enumerate}

\subsection{Derivation} \label{sec:derivation}

\begin{figure}
    \centering
    \includegraphics[width=.6\textwidth]{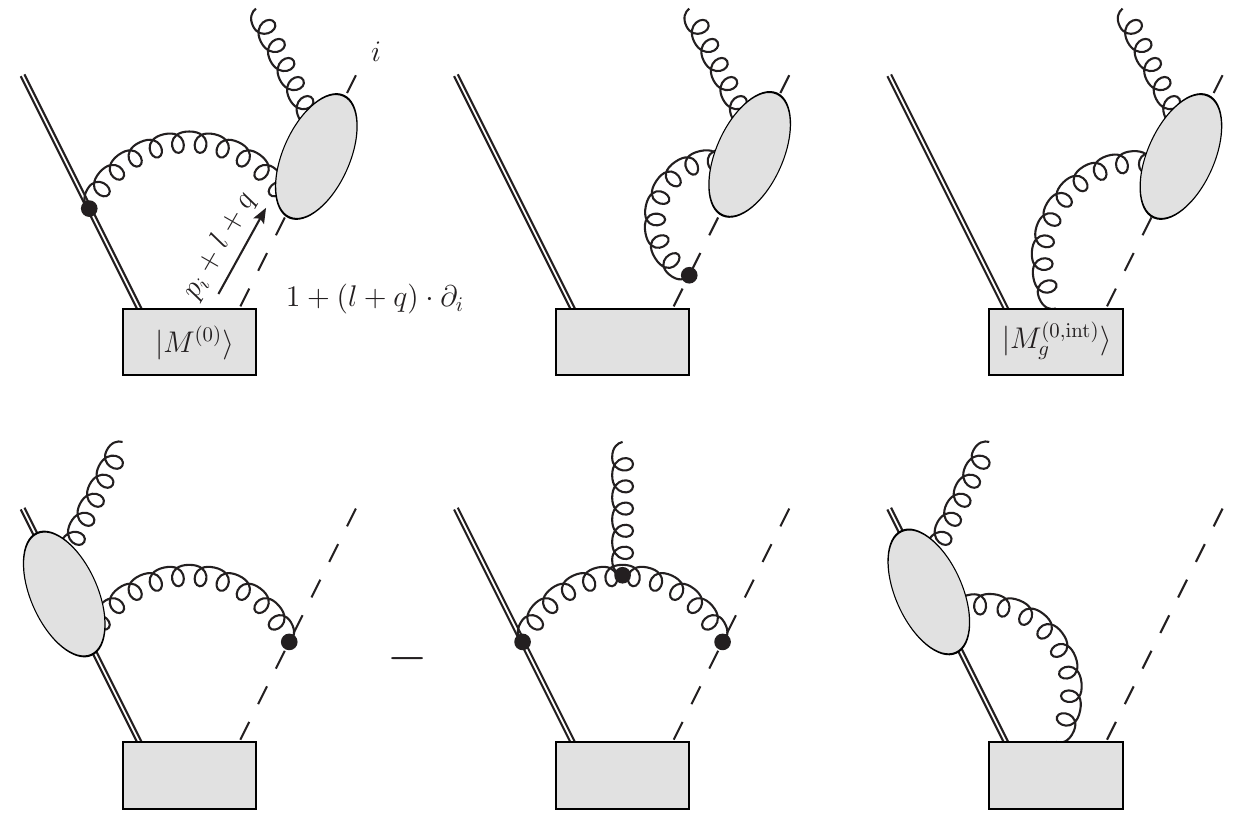} \\[.4cm]
    \includegraphics[width=.36\textwidth]{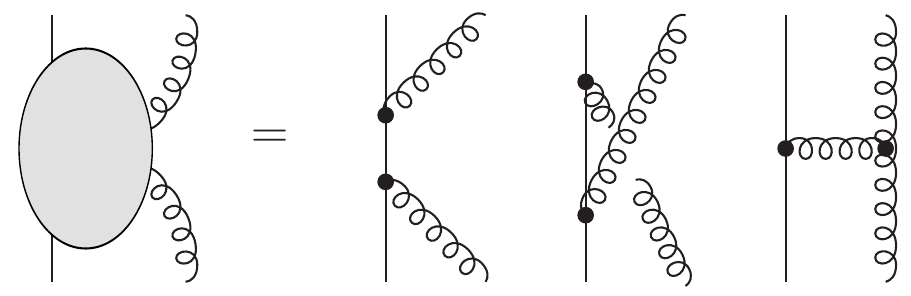} \hspace{1cm}
    \includegraphics[width=.44\textwidth]{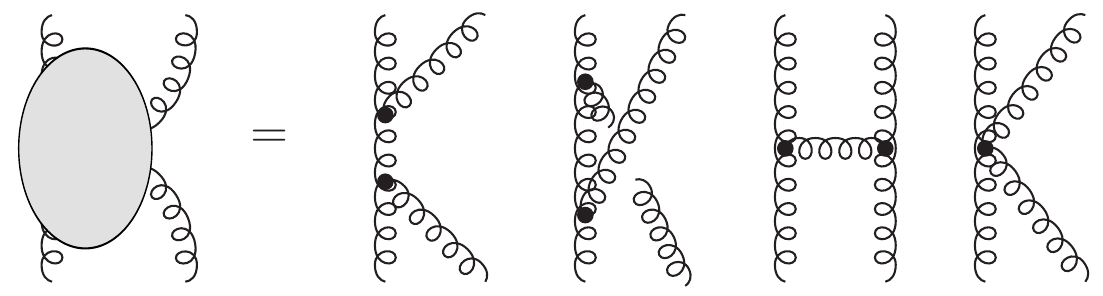}
    \caption{Diagrams included in the calculation of the soft operator. A double line is eikonal, while a dashed line corresponds to a (anti-)quark or a gluon. The shaded rectangle is a process-dependent sum of tree-level diagrams. The eikonal approximation is identical for (anti-)quarks and gluons if expressed through color operators. In this approximation, the quartic gluon vertex does not contribute. It does contribute, however, at subleading order of the soft expansion.}
    \label{fig:diags}
\end{figure}

The result \eqref{eq:S1} for $\mathbf{S}^{(1)}$ is a sum of two contributions of different kind. The first of those is obtained by evaluating the diagrams in Fig.~\ref{fig:diags} after expanding their integrands in the loop momentum and the momentum of the soft gluon. These two momenta are both taken to be formally of the same order, $l^\mu \sim q^\mu \sim \lambda$ with $\lambda$ the soft-expansion parameter. This defines the soft-region contribution within the expansion-by-regions method \cite{Beneke:1997zp} (see also Refs.~\cite{Smirnov:2002pj, Jantzen:2011nz}). Notice that it is necessary to expand the process-dependent part of the amplitude in the soft momenta as demonstrated in the first diagram of Fig.~\ref{fig:diags}. This expansion is responsible for a subset of momentum derivatives present in the soft operator. In case parton $i$ is a gluon, it is necessary to account for the transversality of the process-dependent amplitude. This may generate subleading contributions, because the momentum leaving the process-dependent amplitude on line $i$ is either $p_i+q$, if there are no interactions on line $j$, or $p_i+l+q$. It is this momentum, and not just $p_i$, that contracted with the process-dependent amplitude yields a vanishing contribution.

While constructing the integrand for the soft-region contribution to $\mathbf{S}^{(1)}$ care must be taken of gluon emissions from internal lines of the process-dependent part of the amplitude, as depicted for example in the third diagram of Fig.~\ref{fig:diags}. Following Refs.~\cite{Low:1958sn,Burnett:1967km}, we replace the amplitude with internal-line emissions only, $\ket*{M_g^{(0,\text{int})}}$, by derivatives of the non-radiative amplitude $\ket*{M^{(0)}}$, both defined in Fig.~\ref{fig:diags}\footnote{It may seem that the process-dependent part of the amplitude is not an amplitude itself, because an external spinor or polarization vector is missing on the dashed line. However, if a subleading contribution is considered, as is the case of internal-line emissions, the dashed line is also eikonal.}. To this end, we first insert a gluon in the diagrams of $\ket*{M^{(0)}}$ in all possible ways. We then approximate the resulting radiative amplitude $\ket*{M_g^{(0)}}$ in the soft-gluon limit,
\begin{equation} \label{eq:WardHelper}
    \mathbf{P}(\sigma,c) \ket*{M_g^{0}(\{p_i\},q)} \approx - \sum_i \mathbf{T}_i^c \, \frac{p_i \cdot \epsilon^*}{p_i \cdot q} \big( 1 + q \cdot \tilde{\partial}_i \big) \ket*{M^{(0)}(\{ p_i \})} + \mathbf{P}(\sigma,c) \ket*{M_g^{(0,\text{int}) }(\{p_i\},q = 0)} \; ,
\end{equation}
where $\tilde{\partial}_i$ is a differential operator which vanishes when applied to the spinor or polarization vector on line $i$, while $\tilde{\partial}_i = \partial_i$ otherwise. We have ignored spin effects in \eqref{eq:WardHelper}, because they are represented by an anti-symmetric tensor, which would cancel in the next step anyway. By the Ward identity, the left-hand side of \eqref{eq:WardHelper} vanishes for $\epsilon^* = q$,
\begin{equation}
    0 = - \sum_i \mathbf{T}_i^c \, \big( q \cdot \tilde{\partial}_i \big) \ket*{M^{(0)}(\{ p_i \})} + \mathbf{P}(q,c) \ket*{M_g^{(0,\text{int})}(\{p_i\},q = 0)} \; ,
\end{equation}
where $\mathbf{P}(q,c)$ projects out the scalar polarization by replacing $\epsilon^*$ with $q$. A derivative in $q^\mu$ yields
\begin{equation} \label{eq:Ward}
    \mathbf{P}_g(\mu,c) \ket*{M_g^{(0,\text{int})}} = \sum_{i} \mathbf{T}_i^c \, \tilde{\partial}_i^\mu \ket*{M^{(0)}} \; ,
\end{equation}
where $\mathbf{P}(\mu,c)$ strips the amplitude of the polarization vector and leaves the four-vector index $\mu$ instead. 
The derivatives on the right-hand side of Eq.~\eqref{eq:Ward} complete the set of derivatives occurring in the soft operator.

Remarkably, the soft-operator for a massless (anti-)quark on line $i$ is exactly the same as that for a gluon. This is a consequence of a relationship between double emissions from quark and gluon lines, which is valid at leading and subleading order. In particular, the relevant contributions have the following structure in the Feynman gauge used in the present calculation\footnote{The result for double-gluon emission from a quark line can be obtained with the help of next-to-eikonal Feynman rules presented in Ref.~\cite{Laenen:2010uz}.}.
\begin{equation} \label{eq:blob_quark}
    \vcenter{\hbox{\includegraphics[width=0.15\textwidth]{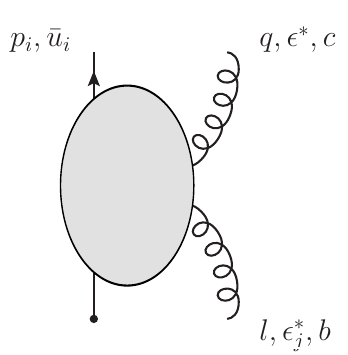}}} = \comm{\mathbf{T}_i^c}{\mathbf{T}_i^b} \Big( \mathcal{C}_{\text{NA}} \, \bar{u}_i + \mathcal{C}_{\text{NA}}^{\mu\nu} \, \bar{u}_i \, \frac{1}{2} \sigma_{\mu\nu} \Big) + \acomm{\mathbf{T}_i^c}{\mathbf{T}_i^b} \Big( \mathcal{C}_{\text{A}} \, \bar{u}_i + \mathcal{C}_{\text{A}}^{\mu\nu} \, \bar{u}_i \, \frac{1}{2} \sigma_{\mu\nu} \Big) + \order{\lambda^0} \; ,
\end{equation}
and
\begin{equation} \label{eq:blob_gluon}
\begin{split}
    \vcenter{\hbox{\includegraphics[width=0.15\textwidth]{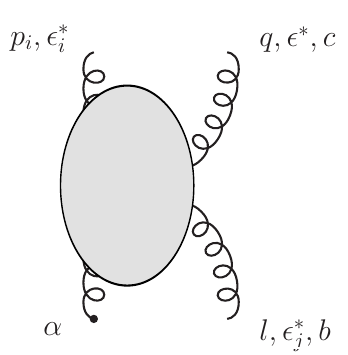}}} &= \quad \comm{\mathbf{T}_i^c}{\mathbf{T}_i^b} \Big( \mathcal{C}_{\text{NA}} \, \epsilon^*_{i\alpha} + \mathcal{C}_{\text{NA}}^{\mu\nu} \, i \big( \epsilon^*_{i\mu}g_{\nu\alpha} - \epsilon^*_{i\nu}g_{\mu\alpha} \big) \Big) \\[-.6cm] &\quad + \acomm{\mathbf{T}_i^c}{\mathbf{T}_i^b} \Big( \mathcal{C}_{\text{A}} \, \epsilon^*_{i\alpha} + \mathcal{C}_{\text{A}}^{\mu\nu} \, i \big( \epsilon^*_{i\mu}g_{\nu\alpha} - \epsilon^*_{i\nu}g_{\mu\alpha} \big) \Big) \\[.2cm]
    &\quad + \mathbf{T}_i^c\mathbf{T}_i^b  \, \frac{\big( \epsilon^*_i \cdot \epsilon^* \big) \, \epsilon^*_{j\alpha}}{2 p_i \cdot q}
    + \mathbf{T}_i^b\mathbf{T}_i^c \, \frac{\big( \epsilon^*_i \cdot \epsilon^*_j \big) \, \epsilon^*_{\alpha}}{2 p_i \cdot l} + \order{\lambda^0} \; ,
\end{split}
\end{equation}
where we have used transversality, $p_i^\alpha = -l^\alpha-q^\alpha$. The common coefficients $\mathcal{C}_{\text{NA},\text{A}}$ and $\mathcal{C}_{\text{NA},\text{A}}^{\mu\nu}$ are linear in $\epsilon^*$ and $\epsilon^*_j$ and depend on the momenta $p_i$, $l$ and $q$. We do not reproduce their values as they are lengthy and not particularly illuminating. Eq.~\eqref{eq:blob_quark} is also valid for a massive quark with exactly the same coefficients. The vector $\epsilon^*_j$ is not assumed transverse to $l$ as it represents the remaining part of the diagram with attachment to line $j$. A propagator is only included on the line with momentum $p_i+l+q$. The terms on the last line in Eq.~\eqref{eq:blob_gluon} yield scaleless integrals. Integrals over $l$ corresponding to diagrams containing the left-hand sides of Eqs.~\eqref{eq:blob_quark}~and~\eqref{eq:blob_gluon} are thus identical if expressed through $J_i - \mathbf{K}_i$ according to Eqs.~\eqref{eq:K}.

For the actual calculation, we have found it practical to perform a Passarino--Veltman reduction followed by an integration-by-parts reduction using \textsc{Kira} \cite{Maierhofer:2017gsa,Klappert:2020nbg}. As a final comment, we point out that the result of integration contains the leading behavior due to $j$ and leading and subleading behavior due to $i$. We have removed the leading behavior due to $j$ in Eq.~\eqref{eq:S1}. It is restored by the summation over $i \neq j$.

The second kind of contributions to $\mathbf{S}^{(1)}$ has a less obvious origin. Consider an amplitude for a process with no massless quarks and only one gluon. Application of the expansion-by-regions method in the soft-gluon limit yields
\begin{equation} \label{eq:ExpansionByRegions}
    \ket*{M_g^{(1)}} \Big|_{\substack{\text{hard} \\ \text{region}}} + \ket*{M_g^{(1)}} \Big|_{\substack{\text{soft} \\ \text{region}}} = \mathbf{S}^{(0)} \, \ket*{M^{(1)}} + \mathbf{S}^{(1)} \, \ket*{M^{(0)}}
    + \text{higher-order soft-expansion terms} \; ,
\end{equation}
where the hard-region contribution is obtained by expanding diagrams in the soft-gluon momentum while keeping the dependence on the loop momentum exact. This expression suggests a one-to-one correspondence between the two terms on each side after appropriate truncation of the expansion. It turns out, however, that there is mismatch
\begin{equation} \label{eq:mismatch}
    \ket*{M_g^{(1)}} \Big|_{\substack{\text{hard} \\ \text{region}}} \neq \mathbf{S}^{(0)} \, \ket*{M^{(1)}}
    + \text{higher-order soft-expansion terms} \; .
\end{equation}

\begin{figure}
    \centering
    \includegraphics[width=.6\textwidth]{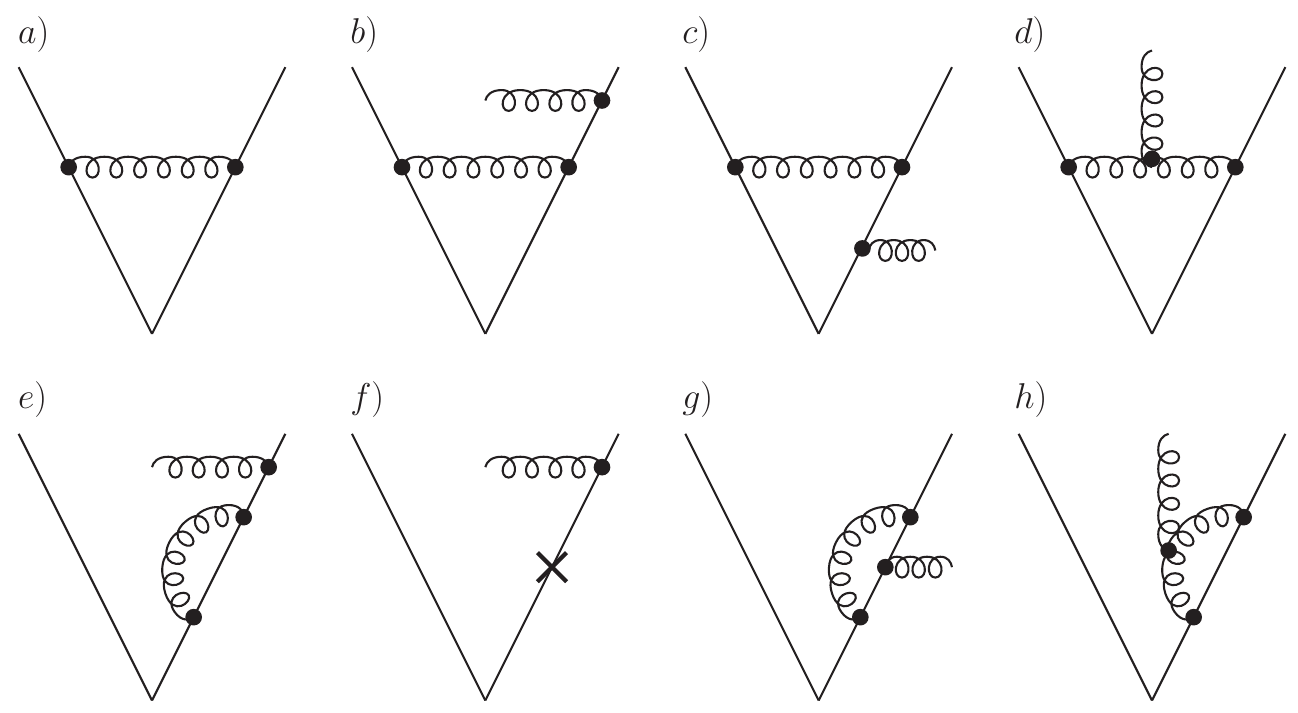}
    \caption{Diagrams for $\gamma^* \to q\bar{q}$ (a)) and $\gamma^* \to q\bar{q}g$ (b)-h)). The cross in diagram f) represents a mass-renormalization counterterm.}
    \label{fig:qqg}
\end{figure}

Let us consider the process $ \gamma^* \to q \bar{q} g$ with a massive quark in order to elucidate the origin of the mismatch. The relevant one-loop diagrams are depicted in Fig.~\ref{fig:qqg}. We omit the LSZ normalization factors as they are the same on both sides of \eqref{eq:mismatch}. $\ket*{M_g^{(1)}} \big|_{\text{hard region}}$ is obtained by expanding the integrands of diagrams b)-h) in $q^\mu \sim \lambda$. Diagrams c) and d) contribute at $\order{\lambda^0}$, b), g) and h) at $\order{1/\lambda}$, and e) and f) at $\order{1/\lambda^2}$. The validity of the leading-power soft factorization requires the sum of e), f), g) and h) to only contribute at $\order{\lambda^0}$. Indeed, the leading, $\order{1/\lambda}$, behavior of the amplitude for $\gamma^* \to q\bar{q}g$ is obtained by multiplying a) with eikonal factors which results in the leading behavior of b) without any further contributions. Let $F_a$ denote the value of diagram a). Similarly, let $F_b$, $F_{cd}$ and $F_{e\dots h}$ denote the values of diagram b), the sum of c) and d), and the sum of e)-h), each expanded to $\order{\lambda^0}$. With this notation
\begin{equation}
    \ket*{M^{(0)}} = F_a \; , \qquad \ket*{M_g^{(1)}} \Big|_{\substack{\text{hard} \\ \text{region}}} = F_b + F_{cd} + F_{e\dots h} + \order{\lambda} \; .
\end{equation}
It follows from the definition \eqref{eq:LBK} of the tree-level soft operator that
\begin{equation} \label{eq:Fb}
    F_b = \Big( \mathbf{S}^{(0)} - \mathbf{S}^{(0,\text{Ward})}\Big) \, F_a \; , \qquad
    \mathbf{S}^{(0,\text{Ward})} \equiv \sum_{i} \mathbf{T}_i^c \, \epsilon^* \cdot \tilde{\partial}_i \; .
\end{equation}
In order to obtain the rest of the contributions, we exploit the Ward identity
\begin{equation}
    \Big[ F_b + F_{cd} + F_{e\dots h} \Big]_{\epsilon^* = q} = 0 \; .
\end{equation}
Substitution of Eq.~\eqref{eq:Fb} yields
\begin{equation}
    \Big[ \mathbf{S}^{(0,\text{Ward})} \Big]_{\epsilon^* = q} \, F_a = \Big[ F_{cd} + F_{e\dots h} \Big]_{\epsilon^* = q} \; . 
\end{equation}
Since both $\mathbf{S}^{(0,\text{Ward})}$ and $F_{cd}$ are independent of $q$, there is
\begin{equation}
    \mathbf{S}^{(0,\text{Ward})} \, F_a = F_{cd} + \epsilon^* \cdot \pdv{q} \Big[ F_{e\dots h} \Big]_{\epsilon^* = q} \; . 
\end{equation}
In consequence
\begin{equation} \label{eq:correction}
    \ket*{M_g^{(1)}} \Big|_{\substack{\text{hard} \\ \text{region}}} = \mathbf{S}^{(0)} \, \ket*{M^{(1)}} + \bigg( F_{e\dots h} - \epsilon^* \cdot \pdv{q} \Big[ F_{e\dots h} \Big]_{\epsilon^* = q} \bigg) + \order{\lambda} \; .
\end{equation}
Although we have obtained an expression for the mismatch \eqref{eq:mismatch} using a specific example, it should be clear at this point that the result only depends on external line diagrams and is thus general. An explicit calculation yields the second contribution to $\mathbf{S}^{(1)}$
\begin{multline} \label{eq:chromomagnetic}
    \sum_i \mathbf{T}_i^c \, \frac{1-\epsilon}{1-2\epsilon} \Big( \big( 1 - 2 \epsilon^2 \big) C_A + 2 \epsilon \big( 1 + 2 \epsilon \big) C_F \Big) \, \frac{A_0(m_i^2)}{m_i^2} \, \bigg( g^{\mu\nu} - \frac{2 p_i^\mu p_i^\nu}{m_i^2} \bigg) \, \frac{F_{\mu\rho}}{s_{iq}} \, \mathbf{K}_{i \, \nu}{}^\rho \\[.2cm]
    = \sum_i \mathbf{T}_i^c \, \bigg[ \bigg( \frac{1}{\epsilon} + \ln \bigg( \frac{\mu^2}{m_i^2} \bigg) + 2 \bigg) C_A + 2  C_F + \order{\epsilon} \bigg] \, \bigg( g^{\mu\nu} - \frac{2 p_i^\mu p_i^\nu}{m_i^2} \bigg) \, \frac{F_{\mu\rho}}{s_{iq}} \, \mathbf{K}_{i \, \nu}{}^\rho\; .
\end{multline}
The coefficient of $C_F$ in $F_{e\dots h}$ satisfies the Ward identity on its own and the derivative contribution in Eq.~\eqref{eq:correction} vanishes. On the other hand, the correct value of the coefficient of $C_A$ requires a non-vanishing derivative contribution. The term proportional to $C_A$ has been included in the coefficient $a_i$ of $\mathbf{S}^{(\text{NA})}_{ij}$ (see Eq.~\eqref{eq:ai}), because it is required to cancel a spurious divergence that would otherwise be present there. The term proportional to $C_F$ has been kept separate in Eq.~\eqref{eq:S1}.

The fact that there are non-trivial hard-region contributions in subleading-soft factorization in QED starting already at one-loop has been pointed out in Refs.~\cite{Engel:2022kde,Kollatzsch:2022bqa}\footnote{The $C_F$ term at $\order{\epsilon^0}$ in Eq.~\eqref{eq:chromomagnetic} matches the right-hand side of Eq.~(28) in Ref.~\cite{Kollatzsch:2022bqa} with appropriate charges.}. Later, in Ref.~\cite{Engel:2023ifn}, these effects have been related to matching coefficients in the abelian version of HQET (Heavy Quark Effective Theory). In HQET, our result \eqref{eq:chromomagnetic} corresponds to corrections to the chromomagnetic operator which are known to three-loop order \cite{Grozin:2007fh}\footnote{After renormalization in the \msbar scheme, \eqref{eq:chromomagnetic} is equal to the one-loop contribution in Eq.~(14) of Ref.~\cite{Grozin:2007fh}.}. 

\subsection{Numerical example}

\begin{figure}
    \centering
    \includegraphics[width=.4\textwidth]{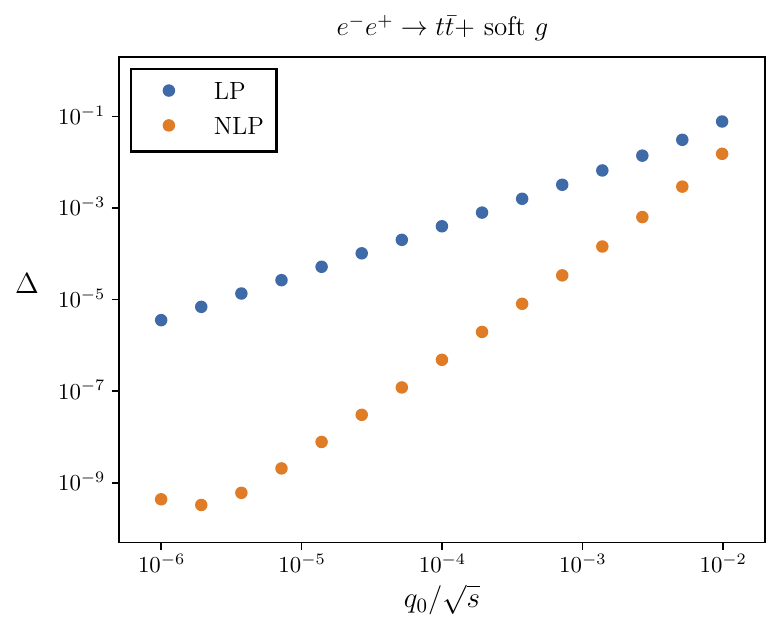} \hspace{1cm}
    \includegraphics[width=.4\textwidth]{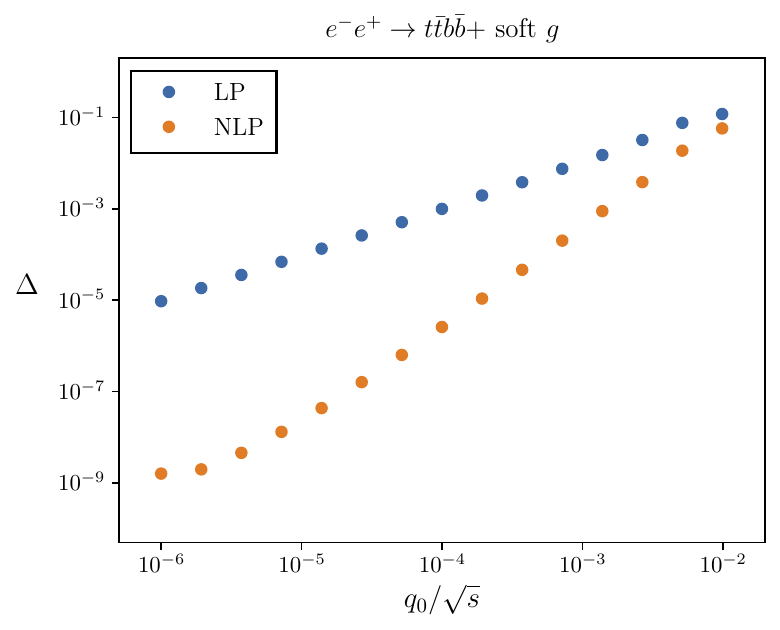}
    \caption{Relative error $\Delta_{\text{LP/NLP}}$ of the one-loop soft approximation to leading power (LP) and subleading power (NLP). The energy, $q_0$, of the soft gluon is normalised to the centre-of-mass energy, $\sqrt{s}$, of the process. The apparent breakdown of the approximation at low soft-gluon energies is due to the limited numerical precision of the one-loop integrals in \textsc{OneLOop} which impacts the result for the $(n+1)$-particle amplitudes.}
    \label{fig:check}
\end{figure}

In order to illustrate the usage of Eq.~\eqref{eq:S1}, we perform a numerical study for two processes: 1) $e^+e^- \to t\bar{t}g$, which is the process discussed at the end of Section~\ref{sec:derivation}; 2) $e^+e^- \to b\bar{b}t\bar{t}g$, which involves quarks with different masses. As in Ref.~\cite{Czakon:2023tld}, the one-loop $n$-particle amplitudes $\ket*{M^{(1)}}$ as well as their derivatives are calculated with \textsc{Recola}~\cite{Actis:2016mpe, Denner:2017wsf} linked to \textsc{Collier}~\cite{Denner:2016kdg, Denner:2002ii, Denner:2005nn, Denner:2010tr} for the evaluation of tensor and scalar one-loop integrals. For the evaluation of the one-loop $(n+1)$-particle amplitudes, $\ket*{M_g^{(1)}}$, we instead link \textsc{Recola} to \textsc{CutTools}~\cite{Ossola:2007ax} for tensor reduction and \textsc{OneLOop}~\cite{vanHameren:2009dr,vanHameren:2010cp} for the evaluation of scalar integrals at quadruple precision.
The quality of the approximations is demonstrated with the help of
\begin{equation} \label{eq:DeltaNLP}
    \Delta_\text{LP/NLP} \equiv \frac{1}{N} \sum_{\substack{\text{singular} \\ \text{colour flows $\{c\}$} \\ \text{helicities $\{\sigma\}$}}} \left|\frac{\left[\braket*{\{c,\sigma\}}{M_{g}^{(1)}} - \braket*{\{c,\sigma\}}{M^{(1)}_g}_\text{LP/NLP}\right]_{\mathcal{O}(\epsilon^0)}}{\left[\braket*{\{c,\sigma\}}{M_{g}^{(1)}}\right]_{\mathcal{O}(\epsilon^0)}}\right| \; ,
\end{equation}
where LP (leading power) stands for soft expansion up to $\order{1/\lambda}$, while NLP (next-to-leading power) up to $\order{\lambda^0}$. The sum runs over all colour-flow and helicity configurations for which the amplitude has a soft singularity. The number of such configurations is denoted by $N$. The quantity $\Delta_\text{LP/NLP}$ is sensitive not only to color effects, but also to spin effects. It is thus better suited than a test on matrix elements summed over polarizations, which are not sensitive to the majority of spin effects. The  results of numerical evaluations are presented in Fig.~\ref{fig:check} and clearly confirm the correctness of the formulae at least for massive quarks and to $\order{\epsilon^0}$.

\section{Future directions}

The formulae that we have provided in this publication and in \cite{Czakon:2023tld} are sufficient to approximate one-loop QCD amplitudes up to the first non-singular term of the soft expansion. Although we have only considered massive quarks, the result would be very similar with massive scalars. In that case, one would need to modify the single-line contributions described in Section~\ref{sec:derivation} and given for massive (anti-)quarks in Eq.~\eqref{eq:chromomagnetic}. In practice, our results can be used to replace the exact amplitude as long as the energy of the gluon is substantially smaller than any other momentum or mass. Clearly, future work should be directed at the emission of two or more soft gluons at one-loop, as well as at the emission of a single gluon beyond the one-loop level. Nevertheless, there remains another problem much closer to the results we have presented: what happens when the momentum of the gluon is much smaller than any other three-momenta, but comparable to the mass of a quark? Surely, this must be described by a formula like Eq.~\eqref{eq:S1}, but the jet operators should now be non-trivial functions of the energy-to-mass ratio. This case has been studied in QED very recently \cite{vanBijleveld:2025ekz}. The determination of the relevant functions will be our next objective.

Finally, let us point out that our factorization formula \eqref{eq:OneLoopLBK} is written in a form which is not easily comparable to the operator formalism of SCET or direct diagrammatic next-to-eikonal exponentiation. This might make it more difficult to apply in an existing resummation framework. On the other hand, our general results are simpler and more compact than partial results derived previously with other methods. It seems that formalisms aiming at resummation make some simplifications less obvious. For this reason, we believe that our presentation can be a valuable starting point of further studies and might lead to novel insights, perhaps even improvements of existing resummation frameworks.

\begin{acknowledgments}
This work was supported by the Deutsche Forschungsgemeinschaft (DFG) under grant 396021762 - TRR 257: Particle Physics Phenomenology after the Higgs Discovery, and grant 400140256 - GRK 2497: The Physics of the Heaviest Particles at the LHC. Diagrams were drawn using \textsc{JaxoDraw}~\cite{Vermaseren:1994je, Collins:2016aya,Binosi:2003yf}.
\end{acknowledgments}

\appendix

\section{Subleading collinear limit of the tree-level $g \to q\bar{q}$ splitting} \label{app:gqq}

Let us consider a process with $n+1$ external particles with particle $i$ a quark and particle $n+1$ an anti-quark. The kinematics of the process is defined by momenta $\{ k_j \}_{j=1}^{n+1}$. The collinear limit under study corresponds to $l_\perp \to 0$ in the Sudakov parameterization
\begin{align}
    &k_{n+1} \equiv x p_i + l_\perp - \frac{l_\perp^2}{2x} \frac{q}{p_i \cdot q} \; , & \text{with} & \qquad l_\perp \cdot p_i = l_\perp \cdot q = q^2 = 0 \; , \\[.2cm]
    & k_i \equiv (1-x) p_i - l_\perp - \frac{l_\perp^2}{2(1-x)} \frac{q}{p_i \cdot q} \; , & \text{and}
    & \qquad k_j \equiv p_j + \order{l_\perp^2} \; , \qquad j \neq i, j \leq n \; .
\end{align}
Using a simple argument based on the fact that the tree-level amplitude for the process is a rational function of $x$ up to the factor due to the external quark/anti-quark spinors, Ref.~\cite{Czakon:2023tld} asserts that\footnote{Notice the slight rearrangement of terms compared to Eq.~(5.5) of Ref.~\cite{Czakon:2023tld}. It implies a different definition of $\ket*{C^{(0)}}$ and is crucial for the neat result presented in Eq.~\eqref{eq:C0}.}
\begin{multline} \label{eq:collinear}
    \ket*{M^{(0)}_{\bar{q}}(\{k_i\}_{i=1}^{n+1})} = 
    \mathbf{Split}^{(0)}_{q\bar{q} \, \leftarrow \, g}(k_i,k_{n+1},p_i) \ket*{M^{(0)}(\{p_j\})} \\[.2cm]
    + \sqrt{x(1-x)} \bigg( \frac{1}{x} \ket*{S^{(0)}} + \frac{1}{1-x} \ket*{\bar{S}^{(0)}} + \sum_I \frac{1}{x_I - x} \ket*{R^{(0)}_I} + \ket*{C^{(0)}} \bigg) + \order{l_\perp} \; .
\end{multline}
The leading behavior in the collinear limit is encapsulated in the splitting function
\begin{equation} \label{eq:SplitQQbarG}
     \mel{c_1,c_2;\sigma_1,\sigma_2}{\mathbf{Split}^{(0)}_{q\bar{q} \, \leftarrow \, g}(k_1,k_2,k)}{c;\sigma} = -\frac{1}{2 \, k_1 \cdot k_2} \, T^{c}_{c_1 c_2} \, \bar{u}(k_1,\sigma_1) \, \slashed{\epsilon}(k,\sigma) \, v(k_2,\sigma_2) \; .
     \end{equation}
This splitting function is just the value of the tree-level diagram for $g \to q\bar{q}$ with an off-shell gluon including the gluon propagator. The remaining terms in formula \eqref{eq:collinear} are given by expressions which are easiest to understand in pictorial form. For the leading term, we write
\begin{equation}
    \mathbf{Split}^{(0)}_{q\bar{q} \, \leftarrow \, g}(k_i,k_{n+1},p_i) \ket*{M^{(0)}(\{p_j\})} = \vcenter{\hbox{\includegraphics[width=0.3\textwidth]{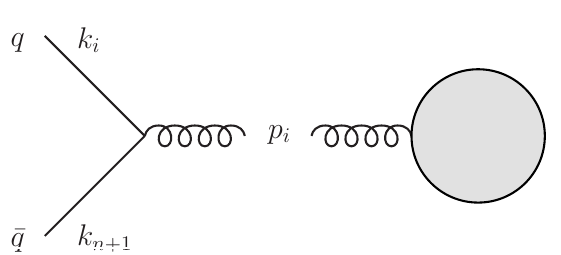}}} \; .
\end{equation}
The shaded circle here and in subsequent equations is an on-shell amplitude for momenta at the limit $l_\perp = 0$ with omitted external particles whenever possible. Because this is an amplitude, a factor $-i$ is always removed. This is particularly important in order to obtain the correct value in Eq.~\eqref{eq:RI}, which contains a product of two amplitudes. Summation over the color and polarization of the intermediate state is implied. The momentum of the intermediate state should be understood as the outgoing momentum in the amplitude on the right.
The subsequent terms are obtained by taking suitable limits in $x$ of the occurring diagrams after removing the leading contribution. For instance, the limit $x \to 0$ yields
\begin{equation}
    \ket*{S^{(0)}} = \sum_{\substack{j\neq i \\ j \leq n}} \quad
    \vcenter{\hbox{\includegraphics[width=0.3\textwidth]{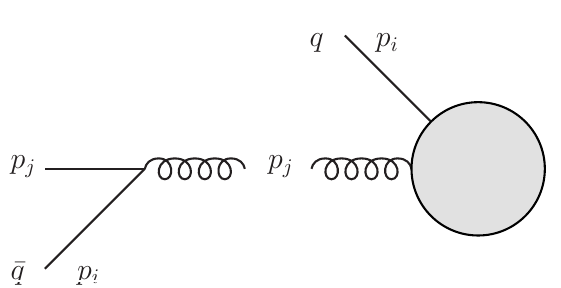}}}
    \quad \text{or} \quad
    \vcenter{\hbox{\includegraphics[width=0.3\textwidth]{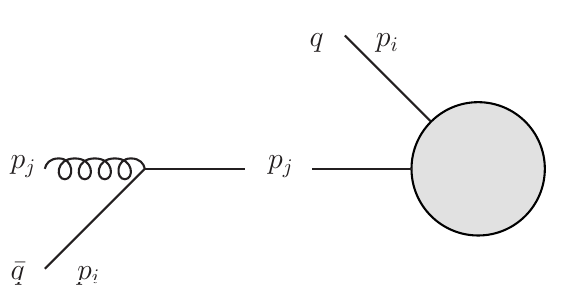}}} \; ,
\end{equation}
where the splitting functions encode the effect of a soft anti-quark. Similarly, the leading asymptotics in the limit $x \to 1$ is due to a soft quark
\begin{equation}
    \ket*{\bar{S}^{(0)}} = \sum_{\substack{j\neq i \\ j \leq n}} \quad
    \vcenter{\hbox{\includegraphics[width=0.3\textwidth]{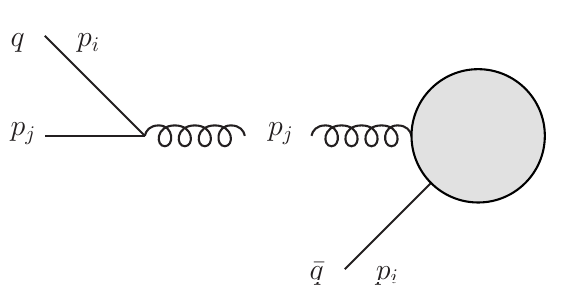}}}
    \quad \text{or} \quad
    \vcenter{\hbox{\includegraphics[width=0.3\textwidth]{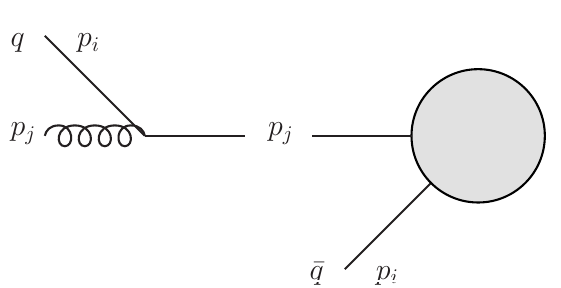}}} \; .
\end{equation}
The tree-level amplitude of the $(n+1)$-parton process exhibits poles at kinematic configurations that allow for on-shell intermediate states. Hence
\begin{equation} \label{eq:RI}
    \ket*{R^{(0)}_I} = \frac{1}{2 p_i \cdot p_I} \frac{1}{\sqrt{x_I(1-x_I)}}
    \vcenter{\hbox{\includegraphics[width=0.4\textwidth]{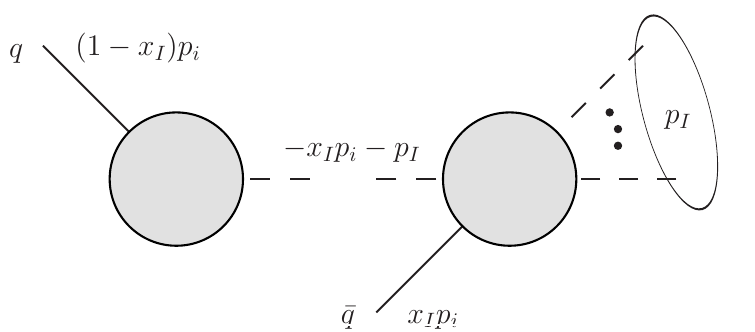}}} \; , \qquad
    x_I = \frac{m_I^2 - p_I^2}{2 p_i \cdot p_I} \; .
\end{equation}
Here, $I$ is a subset of at most $n-3$ of the $n$ particles excluding $i$, so that the result contains two legitimate amplitudes (not splitting functions). The mass of the intermediate state has been denoted with $m_I$, and $p_I$ is the sum of the momenta of the particles in $I$. The dashed lines represent (anti-)quarks, gluons and possibly other particles, if the theory is more general than just QCD.

We do not employ the subleading soft asymptotics to obtain the last contribution, $\ket*{C^{(0)}}$, as was done in Ref.~\cite{Czakon:2023tld} for the other splittings. Instead, we notice that this term corresponds to the limit $|x| \to \infty$. This limit "eikonalises" the collinear quarks, because it effectively makes them infinitely "harder" than any other particle in the process. Hence, the result is simply given by a straight eikonal line with multiple gluons attached as is typical of high energy scattering 
\begin{equation} \label{eq:C0}
    \ket*{C^{(0)}} = -2 \delta_{-\sigma_{n+1}\sigma_i}\sum_{}
    \vcenter{\hbox{\includegraphics[width=0.2\textwidth]{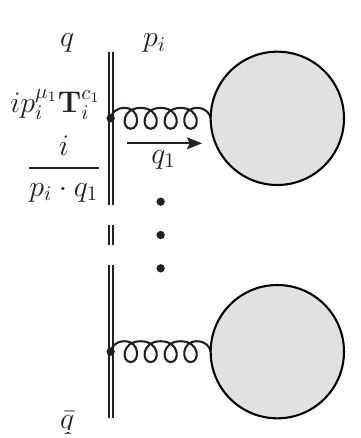}}} \; .
\end{equation}
The sum is taken over all possible tree-level diagrams (disconnected shaded circles) with at least two gluons attaching to the eikonal line. The single-gluon case is already taken care of by the leading collinear asymptotics. There is always one eikonal vertex more than there are eikonal propagators. The result for an outgoing quark-anti-quark pair follows from\footnote{The sign depends on spinor conventions used.}
\begin{equation}
    \bar{u}(p_i,\sigma_i) \gamma^\mu v(p_i,\sigma_{n+1}) = -2p_i^\mu \delta_{-\sigma_{n+1}\sigma_i} \; .
    \end{equation}
The factor of $i$ in the last eikonal vertex due to the spinor contraction comes from the fact that $\ket*{C^{(0)}}$ is an amplitude.

\newpage
\bibliographystyle{JHEP}
\bibliography{main} 

\providecommand{\href}[2]{#2}\begingroup\raggedright\begin{thebibliography}{10}

\bibitem{Czakon:2023tld}
M.~Czakon, F.~Eschment and T.~Schellenberger, \emph{{Subleading effects in
  soft-gluon emission at one-loop in massless QCD}},
  \href{https://doi.org/10.1007/JHEP12(2023)126}{\emph{JHEP} {\bfseries 12}
  (2023) 126} [\href{https://arxiv.org/abs/2307.02286}{{\ttfamily
  2307.02286}}].

\bibitem{Engel:2021ccn}
T.~Engel, A.~Signer and Y.~Ulrich, \emph{{Universal structure of radiative QED
  amplitudes at one loop}},
  \href{https://doi.org/10.1007/JHEP04(2022)097}{\emph{JHEP} {\bfseries 04}
  (2022) 097} [\href{https://arxiv.org/abs/2112.07570}{{\ttfamily
  2112.07570}}].

\bibitem{Engel:2023ifn}
T.~Engel, \emph{{The LBK theorem to all orders}},
  \href{https://doi.org/10.1007/JHEP07(2023)177}{\emph{JHEP} {\bfseries 07}
  (2023) 177} [\href{https://arxiv.org/abs/2304.11689}{{\ttfamily
  2304.11689}}].

\bibitem{Engel:2023rxp}
T.~Engel, \emph{{Multiple soft-photon emission at next-to-leading power to all
  orders}}, \href{https://doi.org/10.1007/JHEP03(2024)004}{\emph{JHEP}
  {\bfseries 03} (2024) 004}
  [\href{https://arxiv.org/abs/2311.17612}{{\ttfamily 2311.17612}}].

\bibitem{Bailhache:2024mck}
R.~Bailhache et~al., \emph{{Anomalous soft photons: Status and perspectives}},
  \href{https://doi.org/10.1016/j.physrep.2024.10.002}{\emph{Phys. Rept.}
  {\bfseries 1097} (2024) 1}
  [\href{https://arxiv.org/abs/2406.17959}{{\ttfamily 2406.17959}}].

\bibitem{Gervais:2017yxv}
H.~Gervais, \emph{{Soft Photon Theorem for High Energy Amplitudes in Yukawa and
  Scalar Theories}},
  \href{https://doi.org/10.1103/PhysRevD.95.125009}{\emph{Phys. Rev. D}
  {\bfseries 95} (2017) 125009}
  [\href{https://arxiv.org/abs/1704.00806}{{\ttfamily 1704.00806}}].

\bibitem{Laenen:2020nrt}
E.~Laenen, J.~Sinninghe~Damst\'e, L.~Vernazza, W.~Waalewijn and L.~Zoppi,
  \emph{{Towards all-order factorization of QED amplitudes at next-to-leading
  power}}, \href{https://doi.org/10.1103/PhysRevD.103.034022}{\emph{Phys. Rev.
  D} {\bfseries 103} (2021) 034022}
  [\href{https://arxiv.org/abs/2008.01736}{{\ttfamily 2008.01736}}].

\bibitem{Bonocore:2015esa}
D.~Bonocore, E.~Laenen, L.~Magnea, S.~Melville, L.~Vernazza and C.D.~White,
  \emph{{A factorization approach to next-to-leading-power threshold
  logarithms}}, \href{https://doi.org/10.1007/JHEP06(2015)008}{\emph{JHEP}
  {\bfseries 06} (2015) 008}
  [\href{https://arxiv.org/abs/1503.05156}{{\ttfamily 1503.05156}}].

\bibitem{Bonocore:2016awd}
D.~Bonocore, E.~Laenen, L.~Magnea, L.~Vernazza and C.D.~White,
  \emph{{Non-abelian factorisation for next-to-leading-power threshold
  logarithms}}, \href{https://doi.org/10.1007/JHEP12(2016)121}{\emph{JHEP}
  {\bfseries 12} (2016) 121}
  [\href{https://arxiv.org/abs/1610.06842}{{\ttfamily 1610.06842}}].

\bibitem{Larkoski:2014bxa}
A.J.~Larkoski, D.~Neill and I.W.~Stewart, \emph{{Soft Theorems from Effective
  Field Theory}}, \href{https://doi.org/10.1007/JHEP06(2015)077}{\emph{JHEP}
  {\bfseries 06} (2015) 077} [\href{https://arxiv.org/abs/1412.3108}{{\ttfamily
  1412.3108}}].

\bibitem{Beneke:2019oqx}
M.~Beneke, A.~Broggio, S.~Jaskiewicz and L.~Vernazza, \emph{{Threshold
  factorization of the Drell-Yan process at next-to-leading power}},
  \href{https://doi.org/10.1007/JHEP07(2020)078}{\emph{JHEP} {\bfseries 07}
  (2020) 078} [\href{https://arxiv.org/abs/1912.01585}{{\ttfamily
  1912.01585}}].

\bibitem{Liu:2021mac}
Z.L.~Liu, M.~Neubert, M.~Schnubel and X.~Wang, \emph{{Radiative quark jet
  function with an external gluon}},
  \href{https://doi.org/10.1007/JHEP02(2022)075}{\emph{JHEP} {\bfseries 02}
  (2022) 075} [\href{https://arxiv.org/abs/2112.00018}{{\ttfamily
  2112.00018}}].

\bibitem{Beneke:2021umj}
M.~Beneke, P.~Hager and R.~Szafron, \emph{{Gravitational soft theorem from
  emergent soft gauge symmetries}},
  \href{https://doi.org/10.1007/JHEP03(2022)199}{\emph{JHEP} {\bfseries 03}
  (2022) 199} [\href{https://arxiv.org/abs/2110.02969}{{\ttfamily
  2110.02969}}].

\bibitem{Beneke:2021aip}
M.~Beneke, P.~Hager and R.~Szafron, \emph{{Soft-collinear gravity beyond the
  leading power}}, \href{https://doi.org/10.1007/JHEP03(2022)080}{\emph{JHEP}
  {\bfseries 03} (2022) 080}
  [\href{https://arxiv.org/abs/2112.04983}{{\ttfamily 2112.04983}}].

\bibitem{Beneke:2022pue}
M.~Beneke, P.~Hager and R.~Szafron, \emph{{Soft-Collinear Gravity and Soft
  Theorems}},  \href{https://arxiv.org/abs/2210.09336}{{\ttfamily 2210.09336}}.

\bibitem{Low:1958sn}
F.E.~Low, \emph{{Bremsstrahlung of very low-energy quanta in elementary
  particle collisions}},
  \href{https://doi.org/10.1103/PhysRev.110.974}{\emph{Phys. Rev.} {\bfseries
  110} (1958) 974}.

\bibitem{Burnett:1967km}
T.H.~Burnett and N.M.~Kroll, \emph{{Extension of the low soft photon theorem}},
  \href{https://doi.org/10.1103/PhysRevLett.20.86}{\emph{Phys. Rev. Lett.}
  {\bfseries 20} (1968) 86}.

\bibitem{DelDuca:1990gz}
V.~Del~Duca, \emph{{High-energy Bremsstrahlung Theorems for Soft Photons}},
  \href{https://doi.org/10.1016/0550-3213(90)90392-Q}{\emph{Nucl. Phys. B}
  {\bfseries 345} (1990) 369}.

\bibitem{vanBeekveld:2023gio}
M.~van Beekveld, A.~Danish, E.~Laenen, S.~Pal, A.~Tripathi and C.D.~White,
  \emph{{Next-to-soft radiation from a different angle}},
  \href{https://doi.org/10.1103/PhysRevD.109.074005}{\emph{Phys. Rev. D}
  {\bfseries 109} (2024) 074005}
  [\href{https://arxiv.org/abs/2308.12850}{{\ttfamily 2308.12850}}].

\bibitem{vanBeekveld:2023liw}
M.~van Beekveld, L.~Vernazza and C.D.~White, \emph{{Exponentiation of soft
  quark effects from the replica trick}},
  \href{https://doi.org/10.1007/JHEP07(2024)109}{\emph{JHEP} {\bfseries 07}
  (2024) 109} [\href{https://arxiv.org/abs/2312.11606}{{\ttfamily
  2312.11606}}].

\bibitem{Balsach:2023ema}
R.~Balsach, D.~Bonocore and A.~Kulesza, \emph{{Soft-photon spectra and the
  Low-Burnett-Kroll theorem}},
  \href{https://doi.org/10.1103/PhysRevD.110.016029}{\emph{Phys. Rev. D}
  {\bfseries 110} (2024) 016029}
  [\href{https://arxiv.org/abs/2312.11386}{{\ttfamily 2312.11386}}].

\bibitem{Pal:2023vec}
S.~Pal and S.~Seth, \emph{{On Higgs+jet production at next-to-leading power
  accuracy}}, \href{https://doi.org/10.1103/PhysRevD.109.114018}{\emph{Phys.
  Rev. D} {\bfseries 109} (2024) 114018}
  [\href{https://arxiv.org/abs/2309.08343}{{\ttfamily 2309.08343}}].

\bibitem{Pal:2024eyr}
S.~Pal and S.~Seth, \emph{{Soft quark effects on H+jet production at NLP
  accuracy}}, \href{https://doi.org/10.1016/j.physletb.2024.139179}{\emph{Phys.
  Lett. B} {\bfseries 860} (2025) 139179}
  [\href{https://arxiv.org/abs/2405.06444}{{\ttfamily 2405.06444}}].

\bibitem{Bierenbaum:2011gg}
I.~Bierenbaum, M.~Czakon and A.~Mitov, \emph{{The singular behavior of one-loop
  massive QCD amplitudes with one external soft gluon}},
  \href{https://doi.org/10.1016/j.nuclphysb.2011.11.002}{\emph{Nucl. Phys. B}
  {\bfseries 856} (2012) 228}
  [\href{https://arxiv.org/abs/1107.4384}{{\ttfamily 1107.4384}}].

\bibitem{Engel:2022kde}
T.~Engel, \emph{{Muon-Electron Scattering at NNLO}}, Ph.D. thesis, Zurich U.,
  2022.
\newblock \href{https://arxiv.org/abs/2209.11110}{{\ttfamily 2209.11110}}.

\bibitem{Kollatzsch:2022bqa}
S.~Kollatzsch and Y.~Ulrich, \emph{{Lepton pair production at NNLO in QED with
  EW effects}},
  \href{https://doi.org/10.21468/SciPostPhys.15.3.104}{\emph{SciPost Phys.}
  {\bfseries 15} (2023) 104}
  [\href{https://arxiv.org/abs/2210.17172}{{\ttfamily 2210.17172}}].

\bibitem{Catani:2000ef}
S.~Catani, S.~Dittmaier and Z.~Trocsanyi, \emph{{One loop singular behavior of
  QCD and SUSY QCD amplitudes with massive partons}},
  \href{https://doi.org/10.1016/S0370-2693(01)00065-X}{\emph{Phys. Lett. B}
  {\bfseries 500} (2001) 149}
  [\href{https://arxiv.org/abs/hep-ph/0011222}{{\ttfamily hep-ph/0011222}}].

\bibitem{Beneke:1997zp}
M.~Beneke and V.A.~Smirnov, \emph{{Asymptotic expansion of Feynman integrals
  near threshold}},
  \href{https://doi.org/10.1016/S0550-3213(98)00138-2}{\emph{Nucl. Phys. B}
  {\bfseries 522} (1998) 321}
  [\href{https://arxiv.org/abs/hep-ph/9711391}{{\ttfamily hep-ph/9711391}}].

\bibitem{Smirnov:2002pj}
V.A.~Smirnov, \emph{{Applied asymptotic expansions in momenta and masses}},
  {\emph{Springer Tracts Mod. Phys.} {\bfseries 177} (2002) 1}.

\bibitem{Jantzen:2011nz}
B.~Jantzen, \emph{{Foundation and generalization of the expansion by regions}},
  \href{https://doi.org/10.1007/JHEP12(2011)076}{\emph{JHEP} {\bfseries 12}
  (2011) 076} [\href{https://arxiv.org/abs/1111.2589}{{\ttfamily 1111.2589}}].

\bibitem{Laenen:2010uz}
E.~Laenen, L.~Magnea, G.~Stavenga and C.D.~White, \emph{{Next-to-Eikonal
  Corrections to Soft Gluon Radiation: A Diagrammatic Approach}},
  \href{https://doi.org/10.1007/JHEP01(2011)141}{\emph{JHEP} {\bfseries 01}
  (2011) 141} [\href{https://arxiv.org/abs/1010.1860}{{\ttfamily 1010.1860}}].

\bibitem{Maierhofer:2017gsa}
P.~Maierh{\"o}fer, J.~Usovitsch and P.~Uwer, \emph{{Kira{\textemdash}A Feynman
  integral reduction program}},
  \href{https://doi.org/10.1016/j.cpc.2018.04.012}{\emph{Comput. Phys. Commun.}
  {\bfseries 230} (2018) 99}
  [\href{https://arxiv.org/abs/1705.05610}{{\ttfamily 1705.05610}}].

\bibitem{Klappert:2020nbg}
J.~Klappert, F.~Lange, P.~Maierh{\"o}fer and J.~Usovitsch, \emph{{Integral
  reduction with Kira 2.0 and finite field methods}},
  \href{https://doi.org/10.1016/j.cpc.2021.108024}{\emph{Comput. Phys. Commun.}
  {\bfseries 266} (2021) 108024}
  [\href{https://arxiv.org/abs/2008.06494}{{\ttfamily 2008.06494}}].

\bibitem{Grozin:2007fh}
A.G.~Grozin, P.~Marquard, J.H.~Piclum and M.~Steinhauser, \emph{{Three-Loop
  Chromomagnetic Interaction in HQET}},
  \href{https://doi.org/10.1016/j.nuclphysb.2007.08.012}{\emph{Nucl. Phys. B}
  {\bfseries 789} (2008) 277}
  [\href{https://arxiv.org/abs/0707.1388}{{\ttfamily 0707.1388}}].

\bibitem{Actis:2016mpe}
S.~Actis, A.~Denner, L.~Hofer, J.-N.~Lang, A.~Scharf and S.~Uccirati,
  \emph{{RECOLA: REcursive Computation of One-Loop Amplitudes}},
  \href{https://doi.org/10.1016/j.cpc.2017.01.004}{\emph{Comput. Phys. Commun.}
  {\bfseries 214} (2017) 140}
  [\href{https://arxiv.org/abs/1605.01090}{{\ttfamily 1605.01090}}].

\bibitem{Denner:2017wsf}
A.~Denner, J.-N.~Lang and S.~Uccirati, \emph{{Recola2: REcursive Computation of
  One-Loop Amplitudes 2}},
  \href{https://doi.org/10.1016/j.cpc.2017.11.013}{\emph{Comput. Phys. Commun.}
  {\bfseries 224} (2018) 346}
  [\href{https://arxiv.org/abs/1711.07388}{{\ttfamily 1711.07388}}].

\bibitem{Denner:2016kdg}
A.~Denner, S.~Dittmaier and L.~Hofer, \emph{{Collier: a fortran-based Complex
  One-Loop LIbrary in Extended Regularizations}},
  \href{https://doi.org/10.1016/j.cpc.2016.10.013}{\emph{Comput. Phys. Commun.}
  {\bfseries 212} (2017) 220}
  [\href{https://arxiv.org/abs/1604.06792}{{\ttfamily 1604.06792}}].

\bibitem{Denner:2002ii}
A.~Denner and S.~Dittmaier, \emph{{Reduction of one loop tensor five point
  integrals}}, \href{https://doi.org/10.1016/S0550-3213(03)00184-6}{\emph{Nucl.
  Phys. B} {\bfseries 658} (2003) 175}
  [\href{https://arxiv.org/abs/hep-ph/0212259}{{\ttfamily hep-ph/0212259}}].

\bibitem{Denner:2005nn}
A.~Denner and S.~Dittmaier, \emph{{Reduction schemes for one-loop tensor
  integrals}},
  \href{https://doi.org/10.1016/j.nuclphysb.2005.11.007}{\emph{Nucl. Phys. B}
  {\bfseries 734} (2006) 62}
  [\href{https://arxiv.org/abs/hep-ph/0509141}{{\ttfamily hep-ph/0509141}}].

\bibitem{Denner:2010tr}
A.~Denner and S.~Dittmaier, \emph{{Scalar one-loop 4-point integrals}},
  \href{https://doi.org/10.1016/j.nuclphysb.2010.11.002}{\emph{Nucl. Phys. B}
  {\bfseries 844} (2011) 199}
  [\href{https://arxiv.org/abs/1005.2076}{{\ttfamily 1005.2076}}].

\bibitem{Ossola:2007ax}
G.~Ossola, C.G.~Papadopoulos and R.~Pittau, \emph{{CutTools: A Program
  implementing the OPP reduction method to compute one-loop amplitudes}},
  \href{https://doi.org/10.1088/1126-6708/2008/03/042}{\emph{JHEP} {\bfseries
  03} (2008) 042} [\href{https://arxiv.org/abs/0711.3596}{{\ttfamily
  0711.3596}}].

\bibitem{vanHameren:2009dr}
A.~van Hameren, C.G.~Papadopoulos and R.~Pittau, \emph{{Automated one-loop
  calculations: A Proof of concept}},
  \href{https://doi.org/10.1088/1126-6708/2009/09/106}{\emph{JHEP} {\bfseries
  09} (2009) 106} [\href{https://arxiv.org/abs/0903.4665}{{\ttfamily
  0903.4665}}].

\bibitem{vanHameren:2010cp}
A.~van Hameren, \emph{{OneLOop: For the evaluation of one-loop scalar
  functions}}, \href{https://doi.org/10.1016/j.cpc.2011.06.011}{\emph{Comput.
  Phys. Commun.} {\bfseries 182} (2011) 2427}
  [\href{https://arxiv.org/abs/1007.4716}{{\ttfamily 1007.4716}}].

\bibitem{vanBijleveld:2025ekz}
R.~van Bijleveld, E.~Laenen, C.~Marinissen, L.~Vernazza and G.~Wang,
  \emph{{Next-to-leading power jet functions in the small-mass limit in QED}},
  \href{https://doi.org/10.1007/JHEP07(2025)257}{\emph{JHEP} {\bfseries 07}
  (2025) 257} [\href{https://arxiv.org/abs/2503.10810}{{\ttfamily
  2503.10810}}].

\bibitem{Vermaseren:1994je}
J.A.M.~Vermaseren, \emph{{Axodraw}},
  \href{https://doi.org/10.1016/0010-4655(94)90034-5}{\emph{Comput. Phys.
  Commun.} {\bfseries 83} (1994) 45}.

\bibitem{Collins:2016aya}
J.C.~Collins and J.A.M.~Vermaseren, \emph{{Axodraw Version 2}},
  \href{https://arxiv.org/abs/1606.01177}{{\ttfamily 1606.01177}}.

\bibitem{Binosi:2003yf}
D.~Binosi and L.~Theussl, \emph{{JaxoDraw: A Graphical user interface for
  drawing Feynman diagrams}},
  \href{https://doi.org/10.1016/j.cpc.2004.05.001}{\emph{Comput. Phys. Commun.}
  {\bfseries 161} (2004) 76}
  [\href{https://arxiv.org/abs/hep-ph/0309015}{{\ttfamily hep-ph/0309015}}].

\end{thebibliography}\endgroup

\end{document}